\documentclass[twocolumn]{autart}    

\usepackage{graphicx}          
                           
\usepackage{amsmath,amssymb}
\usepackage{tikz}
\usepackage{epstopdf}
\usepackage{pgfplots}
\usepackage{xcolor}
\usepackage{caption}
\usepackage{subcaption}
\usepackage{algpseudocode}
\usepackage{algorithm}
\usetikzlibrary{decorations.shapes,shapes}
\allowdisplaybreaks  
\newcommand{\until}[2]{\, U_{[{#1},{#2}]} \, }
\newcommand{\untild}[2]{\, U_{[{#1},{#2}]} \, }
\newcommand{\rs}[1]{\rho^{#1}(\boldsymbol{x},k)}
\newcommand{\rsxi}[1]{\rho_{\boldsymbol{\zeta}}^{#1}(\boldsymbol{x},k)}
\newcommand{\rsxis}[1]{\rho_{soft}^{#1}(\boldsymbol{x},k)}
\newcommand{\rsxie}[1]{\rho_\prime^{#1}(\boldsymbol{x},k)}
\newcommand{\rsxiz}[1]{\rho_{\prime\prime}^{#1}(\boldsymbol{x},k)}
\newcommand{\rss}[2]{\rho^{#1}(\boldsymbol{x},{#2})}
\newcommand{\A}[3]{\mathcal{A}^{#1}(\boldsymbol{#2},#3) }
\newcommand{\Axi}[3]{\mathcal{A}_{\boldsymbol{\zeta}}^{#1}(\boldsymbol{#2},#3) }
\newcommand{\Axie}[3]{\mathcal{A}_\prime^{#1}(\boldsymbol{#2},#3) }
\newcommand{\Axiz}[3]{\mathcal{A}_{\prime\prime}^{#1}(\boldsymbol{#2},#3) }
\newcommand{\As}[3]{\mathcal{A_S}^{#1}(\boldsymbol{#2},#3) }
\newcommand{\Asxi}[3]{\mathcal{A_S}_{\boldsymbol{\zeta}}^{#1}(\boldsymbol{#2},#3) }
\newcommand{\Asxis}[3]{\mathcal{A_S}_{soft}^{#1}(\boldsymbol{#2},#3) }
\newcommand{\Asxie}[3]{\mathcal{A_S}_\prime^{#1}(\boldsymbol{#2},#3) }
\newcommand{\Asxiz}[3]{\mathcal{A_S}_{\prime\prime}^{#1}(\boldsymbol{#2},#3) }

\newcommand{\At}[1]{c_{temp}^{#1}}

\newcommand{\kl}{\Omega(}
\newcommand{\kr}{)}

\newtheorem{problem}{Problem}
\newtheorem{corollary}{Corollary}

\newtheorem{remark}{Remark}
\newtheorem{definition}{Definition}
\newtheorem{example}{Example}

\begin{document}

\begin{frontmatter}

\title{Robust Control for Signal Temporal Logic Specifications using Discrete Average Space Robustness\thanksref{footnoteinfo}} 

\thanks[footnoteinfo]{This work was supported in part by the Swedish Research Council (VR), the European Research Council (ERC), the Swedish Foundation for Strategic Research (SSF), and the Knut and Alice Wallenberg Foundation (KAW).}

\author[KTH]{Lars Lindemann}\ead{llindem@kth.se},    
\author[KTH]{Dimos V. Dimarogonas}\ead{dimos@kth.se}             

\address[KTH]{Department of Automatic Control, School of Electrical Engineering and Computer Science, KTH Royal Institute of Technology, Malvinas v\"ag 10, SE-10044, Stockholm, Sweden}  

\begin{keyword}                           
Formal methods-based control; signal temporal logic; model predictive control; autonomous systems             
\end{keyword}                             

\begin{abstract}                          
Control systems that satisfy temporal logic specifications have become increasingly popular due to their applicability to robotic systems. Existing control methods, however, are computationally demanding, especially when the problem size becomes too large. In this paper, a robust and computationally efficient model predictive control framework for signal temporal logic specifications is proposed. We introduce discrete average space robustness, a novel quantitative semantic for signal temporal logic, that is directly incorporated into the cost function of the model predictive controller. The optimization problem entailed in this framework can be written as a convex quadratic program when no disjunctions are considered and results in a robust satisfaction of the specification. Furthermore, we define the predicate robustness degree as a new robustness notion. Simulations of a multi-agent system subject to complex specifications demonstrate the efficacy of the proposed method.
\end{abstract}

\end{frontmatter}

\section{Introduction}

Formal verification and model checking with specifications in temporal logics \cite{baier} have extensively been studied during the last decade. The dynamical system under consideration is abstracted into a finite state transition system, whereas the specification is translated into a language equivalent b\"uchi automaton. These two representations are then used to systematically check whether or not the specification is satisfied. Conversely, formal methods-based control tries to find control inputs such that the dynamical system satisfies the specification. Formal methods-based control has been investigated with linear temporal logic (LTL) \cite{fainekos2005temporal,guo2015multi,fainekos2009temporal,belta2007symbolic} and metric interval temporal logic (MITL) \cite{nikou1}. LTL and MITL formulas are used to express the system specifications and can be translated into b\"uchi automata \cite{wolper1,maler2}, which allows the calculation of a product automaton with the finite state transition system. Using Dijkstra or a depth-first search algorithm, a suitable discrete run, satisfying the specification, can be found. 

Signal temporal logic (STL) was introduced in \cite{maler1} within the context of monitoring temporal properties of continuous-time signals for continuous and hybrid systems. The authors in \cite{maler1} propose temporal property monitors that check these in STL expressed properties. For instance, for continuous and hybrid control systems it is of great interest to check whether or not the controller satisfies the control specification. STL encompasses a quantitative notion of time and space. The latter property has been used to introduce space robustness (SR) \cite{donze2}, a quantitative semantic stating how robustly a formula is satisfied. SR is a special case of the robust semantics in \cite{fainekos1} for MITL where it was shown that the robust semantics are an under-approximation of the robustness degree. The robustness degree hence indicates how much a signal can be perturbed by noise before changing the truth value of the specification. Consequently, it is possible to use STL to measure how robustly and hence how well a specification is satisfied. This is in contrast to LTL where only a boolean satisfaction is given.  STL was used for control by means of model predictive control (MPC) in \cite{raman1} where SR is incorporated into a mixed integer linear program, which is computationally expensive. Robust extensions of this approach have been reported in \cite{raman2,sadraddini,farahani2015robust}. Additional quantitative semantics have recently been presented. The authors in \cite{donze2} not only define space robustness, but also time robustness and a combination of both. A measure based on the weighted edit distance has been proposed in \cite{jakvsic2016quantitative}, while \cite{brim2013robustness} equips quantitative STL semantics with the possibility to freeze operators. The connection between linear, time-invariant filtering and quantitative semantics has been made in \cite{rodionova2016temporal}. The quantitative semantics in \cite{akazaki2015time}, called averaged STL, extend the semantics in \cite{donze2} by so called averaged operators that average SR over certain time intervals, which naturally leads to space and time robustness. We remark that averaged STL has a similar name to what we define as discrete average space robustness; however, these semantics have fundamental differences. LTL robustness has been considered in \cite{tabuada2015robust}.

The main motivation of this paper is the lack of computationally efficient algorithms for formal methods-based control, especially for systems with fast dynamics where computation time is a critical resource. The contributions of this paper are as follows: we first introduce discrete average space robustness (DASR) as new quantitative semantics over discrete-time signals for STL. DASR semantics are computationally more tractable than SR, but are not an under-approximation of the robustness degree; however, we account for this drawback by including additional constraints into our proposed control strategy. Second, we introduce the notion of the predicate robustness degree and show in an example that DASR, used in an optimization framework, results in a higher predicate robustness degree than SR. Third, a simplified version of DASR is directly incorporated into the cost function of a MPC framework for a fragment of STL formulas. We show that the encapsulated optimization problem can be written as a convex quadratic program. Current approaches as in \cite{sadraddini,raman1,raman2} end up with mixed integer linear programs that become inherently nonconvex if SR is maximized. We directly maximize robustness in a computationally efficient manner. Previous results of our work have been published in \cite{lindemann}. This paper extends \cite{lindemann} by giving geometrical robustness interpretations in terms of hyperplanes and introducing the predicate robustness degree. We motivate why DASR is a good choice for control and explain the $k_1$ calculation for the simplified version of DASR that is missing in \cite{lindemann}. 

Section \ref{sec:preliminaries} introduces notation and preliminaries, while Section \ref{sec:robustness_notions} defines discrete average space robustness. Section \ref{sec:problem_statement} states the formal problem, while Section \ref{sec:control_strategy} presents the proposed solution. Simulations are given in Section~\ref{sec:case_study}, followed by an outlook and conclusions in Section \ref{sec:diskussion_futurework}.

\section{Preliminaries and Notation}
\label{sec:preliminaries}

Scalars are denoted by lowercase, non-bold letters $x$. Column vectors are lowercase, bold letters $\boldsymbol{x}$ and matrices are denoted by uppercase, non-bold letters $X$. The $n$-th row of $X$ will be represented by $X(n,:)$ and similarly $X(:,n)$ represents the $n$-th column of $X$. The $n$-dimensional vector space over the real numbers $\mathbb{R}$ is $\mathbb{R}^n$, whereas $\mathbb{N}$, $\mathbb{R}_{\ge 0}$, and $\mathbb{R}_{> 0}$ are the sets of natural, non-negative, and positive real numbers, respectively. True and false are denoted by $\top$ and $\bot$, while $\otimes$  denotes the Kronecker product; $\boldsymbol{1}_N$ and $\boldsymbol{0}_N$ are column vectors containing $N$ ones and zeros, respectively, whereas $\underbar{0}_{n,n}$ denotes an $n\times n$ matrix consisting of zeros. Let $\|\boldsymbol{x}\|^2_M:=\boldsymbol{x}^TM\boldsymbol{x}$. For two sets $\mathcal{X}$ and $\mathcal{Y}$, the set-valued map $F:\mathcal{X}\rightrightarrows\mathcal{Y}$ maps each $\boldsymbol{x}\in \mathcal{X}$ to a set $F(\boldsymbol{x})\subseteq\mathcal{Y}$.

\subsection{Signals and Systems}

Consider the discrete-time, linear system 
\begin{align}\label{system_discrete}
\boldsymbol{x}(k+1) &= A\boldsymbol{x}(k)+B\boldsymbol{u}(k)
\end{align}
where $A\in \mathbb{R}^{n\times n}$, $B\in \mathbb{R}^{n\times m}$.  Assume that \eqref{system_discrete} has been obtained by sampling a continuous-time system with the sampling function $\tau:\mathbb{N}\to\mathbb{R}_{\ge 0}$ where $\tau(0):=0$. In this paper, periodically sampled systems are considered where $\tau(k):=kT$ with $T$ being the sampling period. For readability reasons, the abbreviation $\tau_k:=\tau(k)$ is used. For a given continuous-time interval $[a,b]$ with $a,b\in\mathbb{R}_{\ge 0}$ and $a\le b$,  define the corresponding discrete-time counterpart as the set
\begin{align}\label{eq:Omega}
\kl a,b \kr:=\{k\in\mathbb{N}|a\le \tau(k) \le b\}.
\end{align} 
\begin{remark}
The definition of $\kl a,b \kr$ allows to impose continuous-time specifications on discrete-time systems. 
\end{remark}

\subsection{Signal Temporal Logic}

Signal temporal logic (STL) consists of predicates $\mu$ that are obtained after evaluation of a predicate function $f:\mathbb{R}^n\to\mathbb{R}$. In particular, for $\boldsymbol{\xi}\in\mathbb{R}^n$, let
\begin{align}\label{eq:predicate}
 \mu:=
 \begin{cases} 
 \top \text{ if } f(\boldsymbol{\xi})\ge0\\
 \bot \text{ if } f(\boldsymbol{\xi})<0.
 \end{cases}
\end{align} 
Let now $\boldsymbol{x}:\mathbb{N}\to\mathbb{R}^n$ be a discrete-time signal and possibly a solution to \eqref{system_discrete}, then $f(\boldsymbol{x}(k))$ determines the truth value of $\mu$ at time $k$. Let $N_\mu$ indicate the number of predicates under consideration and let the set of predicates be $\mathcal{P}:=\{ \mu_1, \mu_2, \cdots, \mu_{N_\mu} \}$; $\mu_i\in \mathcal{P}$ at time $k$ is evaluated as in \eqref{eq:predicate} by $f_i(\boldsymbol{x}(k))$, which is abbreviated by
\begin{align*}
z_i(k):=f_i(\boldsymbol{x}(k)) \;\;\; \forall i\in\{1,2,\cdots, N_\mu\} 
\end{align*} 
for readability reasons. The predicate vector $\boldsymbol{z}(k)$, gathering all predicate functions $z_i(k)$, is then defined as 
\begin{align}\label{pred_nonlinear}
\boldsymbol{z}(k) := 
\begin{bmatrix}
z_1(k)&
\hdots&
z_{N_\mu}(k)
\end{bmatrix}^T.  
\end{align}
If, for all $i\in\{1,\hdots,N_\mu\}$, $z_i(k):=f_i(\boldsymbol{x}(k))$ is affine in $\boldsymbol{x}(k)$,  the predicate functions can be combined with the system in \eqref{system_discrete} to obtain a compact linear form. Define
\begin{align}\label{eq:stacked_pred}
\boldsymbol{z}(k) := 
\begin{bmatrix}
z_1(k)&
\hdots&
z_{N_\mu}(k)
\end{bmatrix}^T:=C\boldsymbol{x}(k)+\boldsymbol{c},
\end{align}
where $C\in\mathbb{R}^{N_{\mu} \times n}$ and $\boldsymbol{c} \in \mathbb{R}^{N_{\mu}}$ encode the predicates under consideration. Using the prediction horizon $N$, let $\boldsymbol{z}_{st}:=\begin{bmatrix}
{\boldsymbol{z}(k_0+1)}^T&
{\boldsymbol{z}(k_0+2)}^T&
\hdots&
{\boldsymbol{z}(k_0+N)}^T
\end{bmatrix}^T$ and $\boldsymbol{u}_{st}:=\begin{bmatrix}
{\boldsymbol{u}(k_0)}^T&
{\boldsymbol{u}(k_0+1)}^T&
\hdots&
{\boldsymbol{u}(k_0+N-1)}^T
\end{bmatrix}^T$ to denote predicates and inputs at different times with respect to the initial time $k_0$, respectively. Inserting the solution $\boldsymbol{x}$ of \eqref{system_discrete} with initial time $k_0$ into \eqref{eq:stacked_pred}, we get
\begin{align}\label{eq:transformation}
\boldsymbol{z}_{st}:=H_1\boldsymbol{x}(k_0)+H_2\boldsymbol{u}_{st}+\boldsymbol{1}_N\otimes \boldsymbol{c},
\end{align}
where $H_1$ and $H_2$ follow recursively by \eqref{system_discrete} and \eqref{eq:stacked_pred}.

Considering $\mu \in \mathcal{P}$, the STL syntax is now given by  
\begin{align*}
\phi \; ::= \; \top \; | \; \mu \; | \; \neg \phi \; | \; \phi \wedge \psi \; | \; \phi  \until{a}{b} \psi\;,
\end{align*}
where  $\phi$, $\psi$ are STL formulas. The STL semantics have been defined over continuous-time signals \cite{maler1,donze2}. Inferring satisfaction of a formula defined over continuous-time signals by considering discrete-time signals has been addressed in \cite[Section~4]{fainekos1} and is possible under some conditions on $\tau$ and $\boldsymbol{x}$. We define the semantics for STL over discrete-time signals by minor changes in the definitions of \cite{maler1,donze2}, hence resembling the discrete definitions in \cite[Section~3]{fainekos1}. The satisfaction relation $(\boldsymbol{x},k)\models \phi$ denotes if the signal $\boldsymbol{x}$ satisfies $\phi$ at time $k$. Note the use of $\Omega(a,b)$ from \eqref{eq:Omega} and recall $\tau_k:=\tau(k)$.

\begin{definition} Given $\mu_i\in \mathcal{P}$, STL formulas $\phi$ and $\psi$, and a signal $\boldsymbol{x}$, the STL semantics \cite{donze2,raman1} are defined as:
\begin{align*}
&(\boldsymbol{x},k) \models \mu_i 				 	&\Leftrightarrow	\;\;\; 	&f_i(\boldsymbol{x}(k))\ge0\\
&(\boldsymbol{x},k) \models \neg\phi 			 	&\Leftrightarrow	\;\;\; 	&\neg((\boldsymbol{x},k) \models \phi)\\
&(\boldsymbol{x},k) \models \phi \wedge \psi 	 	&\Leftrightarrow \;\;\;	
&(\boldsymbol{x},k) \models \phi \wedge (\boldsymbol{x},k) \models \psi\ \\
&(\boldsymbol{x},k) \models \phi \until{a}{b} \psi		&\Leftrightarrow \;\;\;	 	&\exists k_1 \in \kl \tau_k+a,\tau_k+b \kr \text{ s.t. } \\ 
& & &\hspace{-2cm}(\boldsymbol{x},k_1)\models \psi \wedge \forall k_2\in \kl \tau_k,\tau_{k_1} \kr, (\boldsymbol{x},k_2) \models \phi
\end{align*}

\label{def:01} 
\end{definition}
\vspace{-0.5cm}
 Disjunction-, eventually-, and always-operator are derived as $\phi\vee\psi:=\neg(\neg\phi\wedge\neg\psi)$, $F_{[ a,b ]}\phi:=\top  \until{a}{b} \phi$, and $G_{[ a,b ]}\phi := \neg F_{[ a,b ]}\neg\phi$, respectively. The discrete length $h_d^\phi$ of a formula $\phi$ is defined in Definition \ref{def:6} and can be interpreted as the prediction horizon that is sufficient to evaluate $(\boldsymbol{x},k) \models \phi$. For instance, if $h^\phi_d:=5$, then having knowledge of $\boldsymbol{x}(k^\prime)$ for all $k^\prime\in \{ k,k+1,\hdots,k+5 \}$ is sufficient to evaluate $(\boldsymbol{x},k) \models \phi$. 
\begin{definition}{The continuous formula length $h_c^\phi$ is defined as \cite{maler1}:}
$h_c^\mu:= 0$, $h_c^{\neg \phi} := h_c^\phi$, $h_c^{\phi \until{a}{b} \psi} := b+\max(h_c^\phi,h_c^\psi)$, $h_c^{G_{[ a,b ]}\phi}:=h_c^{F_{[ a,b ]}\phi}:=b+h_c^\phi$, $h_c^{\phi \wedge \psi}:=h_c^{\phi \vee \psi} := \max(h_c^\phi,h_c^\psi)$. The discrete formula length is $h_d^\phi:=\max_{k^*\in \kl 0,h^\phi_c \kr} k^*$, i.e., $h_d^\phi$ is the largest element in $\kl 0,h^\phi_c \kr$.
\label{def:6}
\end{definition}

\subsection{Robustness Degree and Space Robustness}

Robustness of a discrete-time signal $\boldsymbol{x}$ with respect to a formula $\phi$ has been defined by the discrete-time robustness degree, robustness degree for short, as how much the signal $\boldsymbol{x}$ can be perturbed before the boolean evaluation of $(\boldsymbol{x},k)\models \phi$ changes its truth value \cite{fainekos1}.  Let $\mathcal{L}_k(\phi):=\{\boldsymbol{x}:\mathbb{N}\to\mathbb{R}^n|(\boldsymbol{x},k)\models\phi\}$ be the set of discrete-time signals that satisfy $\phi$ at time $k$. The closeness of two signals $\boldsymbol{x}$ and $\boldsymbol{x}^*$ is given by $\rho(\boldsymbol{x},\boldsymbol{x}^*):=\sup_{k\in\mathbb{N}} d\big(\boldsymbol{x}(k),\boldsymbol{x}^*(k)\big)$ where $d$ is a metric assigning a distance in $\mathbb{R}^n$; $d$ can be the Euclidean distance $d(\boldsymbol{x}(k),\boldsymbol{x}^*(k)):=\|\boldsymbol{x}(k)-\boldsymbol{x}^*(k)\|$. The distance of $\boldsymbol{x}$ to the set $\mathcal{L}_k(\phi)$ is defined as 
\begin{align*}
\mathcal{\delta}_k(\boldsymbol{x},\phi):=\inf_{\boldsymbol{x}^*\in \text{cl}(\mathcal{L}_k(\phi))}\rho(\boldsymbol{x},\boldsymbol{x}^*),
\end{align*} 
where $\text{cl}(\mathcal{L}_k(\phi))$ denotes the closure of $\mathcal{L}_k(\phi)$. The robustness degree is now given in Definition \ref{def:rd}.

\begin{definition}\label{def:rd}
Given a formula $\phi$ and a signal $\boldsymbol{x}$, the robustness degree at time $k$ is defined as \cite[Definition~23]{fainekos1}:
\begin{align*}
\mathcal{RD}_k(\boldsymbol{x},\phi):=
\begin{cases}
-\delta_k(\boldsymbol{x},\phi) \text{ if } \boldsymbol{x}\notin \mathcal{L}_k(\phi)\\
\delta_k(\boldsymbol{x},\neg\phi) \text{ if } \boldsymbol{x}\in \mathcal{L}_k(\phi)
\end{cases}
\end{align*}
\end{definition}
Furthermore, there are quantitative semantics that state how robustly $\phi$ is satisfied. Such quantitative semantics are given by space robustness (SR)\cite{donze2}, which reasons over continuous-time signals. A modified discrete-time version, denoted by $\rs{\phi}$, is given in Definition \ref{def:2}, which resembles the discrete-time robust semantics in \cite[Definition~26]{fainekos1}. SR is an under-approximation of the robustness degree, i.e., $-\delta_k(\boldsymbol{x},\phi)\le \rs{\phi} \le \delta_k(\boldsymbol{x},\neg \phi)$ \cite[Theorem~28]{fainekos1} so that $|\rs{\phi}|\le |\mathcal{RD}_k(\boldsymbol{x},\phi)|$. Furthermore, $(\boldsymbol{x},k)\models \phi$ if $\rs{\phi}>0$ \cite[Theorem~30]{fainekos1}.
\begin{definition}{Given $\mu_i\in \mathcal{P}$, STL formulas $\phi$ and $\psi$, and a signal $\boldsymbol{x}$, space robustness (SR) is defined as \cite{donze2,raman1}:}
\begin{align*}
\rs{\mu_i} &:= f_i(\boldsymbol{x}(k))=z_i(k)\\
\rs{\neg\phi} &:= 	-\rs{\phi}\\
\rs{\phi \wedge \psi} &:= 	\min(\rs{\phi},\rs{\psi})\\
\rs{\phi \vee \psi} &:= \max(\rs{\phi},\rs{\psi})\\
\rs{\phi \until{a}{b} \psi} &:= \underset{k_1\in \kl \tau_k+a,\tau_k+b \kr }{\max}\min\big(\rss{\psi}{k_1},\\
& \hspace{2.4cm}\underset{k_2\in \kl \tau_k,\tau_{k_1} \kr }{\min}\rss{\phi}{k_2}\big)\\
\rs{F_{[ a,b ]} \phi} &:= \underset{k_1\in \kl \tau_k+a,\tau_k+b \kr}{\max}\rss{\phi}{k_1}\\
\rs{G_{[ a,b ]} \phi} &:= \underset{k_2\in \kl \tau_k+a,\tau_k+b \kr }{\min}\rss{\phi}{k_2}
\end{align*}
\label{def:2}
\end{definition}

\section{Discrete Average Space Robustness and the Predicate Robustness Degree}
\label{sec:robustness_notions}
Discrete average space robustness is defined in Section~\ref{sec:ASR}, while Section~\ref{sec:predicate_robustness} investigates how robustness is affected by changes in $\boldsymbol{z}(k)$.  The results of Section~\ref{sec:predicate_robustness} are then extended in Section \ref{sec:PRD} to define the predicate robustness degree. In the remainder, we consider the formula $\phi$ to be in positive normal form (PNF) \cite[Appendix~A.3]{baier}. For formulas in PNF, negations only occur in front of predicates, e.g., $\neg G_{[ a,b] } \mu_1$ is not in PNF, while $G_{[ a,b ] } \neg \mu_1:=G_{[ a,b ] } \mu_2$ with $\mu_2:=\neg\mu_1$ is in PNF. A formula that is not in PNF can be rewritten in PNF \cite{sadraddini}. Subsequently,  we assume that $\phi$ contains no negations since each negation can be encoded into a new predicate, e.g., $\mu_2:=\neg \mu_1$ for the example above.

\subsection{Discrete Average Space Robustness}
\label{sec:ASR}

Space robustness (Definition \ref{def:2}) uses min-operations to consider the point of weakest satisfaction within a signal. First, define $k_{min}:=\min_{k^*\in \kl \tau_k+a,\tau_k+b \kr} k^*$ and $k_{max}:=\max_{k^*\in \kl \tau_k+b,\tau_k+b\kr} k^*$ as the smallest and largest element of $\kl \tau_k+a,\tau_k+b \kr$, respectively. We now propose novel quantitative semantics $\A{\phi}{x}{k}$ in Definition~\ref{def:44}, which are called discrete average space robustness (DASR) and where instead average satisfaction is considered. An advantage of DASR compared to SR is that min-operations of the always- and until-operators are replaced by linear expressions. This makes DASR computationally more tractable than SR.
\begin{definition}{Given $\mu_i\in \mathcal{P}$, STL formulas $\phi$ and $\psi$, and a signal $\boldsymbol{x}$, discrete average space robustness (DASR) is defined as:}
\begin{align*}
\A{\mu_i}{x}{k} &:= f_i(\boldsymbol{x}(k))=z_i(k)\\
\A{\neg\phi}{x}{k} &:= -\A{\phi}{x}{k}\\
\A{\phi \wedge \psi}{x}{k} &:= \min(\A{\phi}{x}{k},\A{\psi}{x}{k})\\
\A{\phi \vee \psi}{x}{k} &:= \max(\A{\phi}{x}{k},\A{\psi}{x}{k})\\
\A{\phi \untild{a}{b}\psi}{x}{k}&:= \underset{k_1\in\kl \tau_k+a,\tau_k+b \kr }{\max} \frac{1}{2}\biggl( \frac{1}{k_1-k+1} \\
			& \hspace{1cm}\cdot \sum_{k^\prime=k}^{k_1}\A{\phi}{x}{k^\prime}  + \A{\psi}{x}{k_1} \biggr)\\
\A{F_{[ a,b] }\phi}{x}{k}&:=\underset{k_1\in\kl \tau_k+a,\tau_k+b\kr }{\max} \A{\phi}{x}{k_1}\\
\A{G_{[ a,b]}\phi}{x}{k}&:= \frac{1}{k_{max}-k_{min}+1}\sum_{k^\prime=k_{min}}^{k_{max}} \A{\phi}{x}{k^\prime}
\end{align*}
\label{def:44}
\end{definition}
\vspace{-6mm}
To further remove the max-operations of the eventually- and until-operators, $k_1$ can be manually chosen. A method to determine $k_1$ is described in Section \ref{sec:k1}. This results in a simplified version of DASR in Definition \ref{def:55}, which we call discrete simplified average space robustness (DSASR) and denote by $\As{\phi}{x}{k}$. 
\begin{definition}{Given $\mu_i\in \mathcal{P}$, $k_1\in\mathbb{N}$ with $k_1\ge k$, STL formulas $\phi$ and $\psi$, and a signal $\boldsymbol{x}$, discrete simplified average space robustness (DSASR) is defined as:}
\begin{align*}
\As{\mu_i}{x}{k} &:= f_i(\boldsymbol{x}(k))=z_i(k)\\
\As{\neg\phi}{x}{k} &:= -\As{\phi}{x}{k}\\
\As{\phi \wedge \psi}{x}{k} &:= \min(\As{\phi}{x}{k},\As{\psi}{x}{k})\\
\As{\phi \vee \psi}{x}{k} &:= \max(\As{\phi}{x}{k},\As{\psi}{x}{k})\\
\As{\phi \untild{a}{b}\psi}{x}{k}&:=\frac{1}{2} \biggl( \frac{1}{k_1-k+1} \sum_{k^\prime=k}^{k_1}\As{\phi}{x}{k^\prime} \\
					& \hspace{3.5cm}+ \As{\psi}{x}{k_1}\biggr)\\
\As{F_{[ a,b]}\phi}{x}{k}&:= \As{\phi}{x}{k_1}\\
\As{G_{[ a,b]}\phi}{x}{k}&:= \frac{1}{k_{max}-k_{min}+1}\sum_{k^\prime=k_{min}}^{k_{max}} \As{\phi}{x}{k^\prime}
\end{align*}
\label{def:55}
\end{definition}
\vspace{-0.5cm} 
Note that unlike SR, neither DASR nor DSASR satisfy that $(\boldsymbol{x},k)\models \phi$ if $\A{\phi}{x}{k}>0$ or $\As{\phi}{x}{k}>0$. This can be seen by considering $\phi:=G_{[ a,b ]}(x\ge 0)$, where it is possible that, even if $\A{\phi}{x}{k}=\frac{1}{k_{max}-k_{min}+1}\sum_{k^\prime=k_{min}}^{k_{max}}x(k^\prime)>0$, there might exist a $k_2 \in \kl \tau_k+a,\tau_k+b \kr$ s.t. $x(k_2)<0$ and hence $\rss{\phi}{k}=\underset{k_2\in \kl \tau_k+a,\tau_k+b\kr}{\text{min}}x(k_2)<0$. Subsequently, $\A{\phi}{x}{k}>0 \nRightarrow (\boldsymbol{x},k) \models \phi $, whereas $\rs{\phi} > 0 \Rightarrow (\boldsymbol{x},k) \models \phi$. However, in this paper additional constraints will be included in the optimization problem to enforce this property. Another property of DASR and DSASR is the following: it holds that $\mathcal{A}^{G_{[a,b]}\phi}(\boldsymbol{x},k)\neq\mathcal{A}^{\neg F_{[a,b]} \neg \phi}(\boldsymbol{x},k)$ and $\mathcal{A_S}^{G_{[a,b]}\phi}(\boldsymbol{x},k)\neq\mathcal{A_S}^{\neg F_{[a,b]} \neg \phi}(\boldsymbol{x},k)$ although $G_{[a,b]}\phi=\neg F_{[a,b]} \neg \phi$. Furthermore, $\mathcal{A}^{\top \mathcal{U}_{[a,b]} \phi}(\boldsymbol{x},k)\neq\mathcal{A}^{F_{[a,b]} \phi }(\boldsymbol{x},k)$ and $\mathcal{A_S}^{\top \mathcal{U}_{[a,b]}  \phi}(\boldsymbol{x},k)\neq\mathcal{A_S}^{F_{[a,b]} \phi }(\boldsymbol{x},k)$ although $\top \mathcal{U}_{[a,b]}  \phi=F_{[a,b]}\phi$. These cases, however, will not occur in the STL fragment that is considered in Section \ref{sec:problem_statement}. We remark that the goal is to obtain expressive and computationally tractable semantics.
\begin{remark}
Averaged STL was introduced in \cite{akazaki2015time} and is different compared to DASR and DSASR. The averaged always-, eventually-, and until-operators of averaged STL form a weighted time average over $\rs{G_{[ a,b]} \phi}$, $\rs{F_{[ a,b]} \phi}$, and $\rs{\phi \until{a}{b} \psi}$, respectively, hence not removing min- or max-operations and keeping nonlinear, nonsmooth, and nonconvex semantics of the always-, eventually-, and until-operators. This can cause computational burdens in optimization problems. Furthermore, averaged STL results in time and space robustness, while DASR and DSASR only consider a notion of space robustness. Time robustness is a useful measure with the drawback of resulting in even more complex semantics.
\end{remark}

\subsection{Predicate Functions as a Robustness Indicator}
\label{sec:predicate_robustness}

Define a vector $\boldsymbol{\zeta}:=\begin{bmatrix} \zeta_1 & \zeta_2 & \hdots & \zeta_{N_\mu}\end{bmatrix}^T\in\mathbb{R}_{>0}^{N_\mu}$ and consider the possibly nonlinear predicate vector $\boldsymbol{z}(k)$ as in \eqref{pred_nonlinear}. Furthermore, define a modified predicate vector $\boldsymbol{z}_{\boldsymbol{\zeta}}(k):=\boldsymbol{z}(k)+\boldsymbol{\zeta}$. It hence holds that $\boldsymbol{z}(k)<\boldsymbol{z}_{\boldsymbol{\zeta}}(k)$ where $<$ indicates element-wise inequality.  Assume that the formula $\phi$ contains the predicates $\mathcal{P}:=\{\mu_1,\hdots,\mu_{N_\mu}\}$ and that $(\boldsymbol{x},k)\models \phi$ is evaluated with two different predicate vectors $\boldsymbol{z}(k)$ and $\boldsymbol{z}_{\boldsymbol{\zeta}}(k)$. In other words, each predicate $\mu_1,\hdots,\mu_{N_\mu}$ is evaluated as in \eqref{eq:predicate} with the corresponding elements of either $\boldsymbol{z}(k)$ or $\boldsymbol{z}_{\boldsymbol{\zeta}}(k)$. According to Definitions \ref{def:2}, \ref{def:44}, and \ref{def:55}, we define and obtain $\rs{\phi}$, $\A{\phi}{x}{k}$, and $\As{\phi}{x}{k}$ if $\boldsymbol{z}(k)$ is used or $\rsxi{\phi}$, $\Axi{\phi}{x}{k}$, and $\Asxi{\phi}{x}{k}$ if $\boldsymbol{z}_{\boldsymbol{\zeta}}(k)$ is used. 
\begin{corollary}\label{cor1}
Consider a formula $\phi$ in PNF and two predicate vectors $\boldsymbol{z}(k)$ and $\boldsymbol{z}_{\boldsymbol{\zeta}}(k)$ such that $\boldsymbol{z}(k)<\boldsymbol{z}_{\boldsymbol{\zeta}}(k)$. It then holds that $\rs{\phi}<\rsxi{\phi}$, $\A{\phi}{x}{k}<\Axi{\phi}{x}{k}$, and $\As{\phi}{x}{k}<\Asxi{\phi}{x}{k}$. Furthermore, it holds that $\rsxi{\phi}-\rs{\phi}\ge \zeta_{min}$, $\Axi{\phi}{x}{k}-\A{\phi}{x}{k}\ge \zeta_{min}$, and $\Asxi{\phi}{x}{k}-\As{\phi}{x}{k}\ge \zeta_{min}$ where $\zeta_{min}:= \min_{i\in \{1,\hdots,N_\mu \}} \zeta_i$. 
\end{corollary}
\begin{pf}
A formula $\phi$ in PNF excludes the case of negations in Definitions \ref{def:2}, \ref{def:44}, and \ref{def:55}. Due to the inductive definition of SR, DASR, DSASR, and the exclusion of negations, a bigger predicate vector $\boldsymbol{z}(k)$ will lead to bigger functions $\rs{\phi}$, $\A{\phi}{x}{k}$, and $\As{\phi}{x}{k}$. Since $\boldsymbol{z}(k)<\boldsymbol{z}_{\boldsymbol{\zeta}}(k)$ holds element-wise, we can conclude that  $\rs{\phi}<\rsxi{\phi}$, $\A{\phi}{x}{k}<\Axi{\phi}{x}{k}$, and $\As{\phi}{x}{k}<\Asxi{\phi}{x}{k}$. It is then straightforward to show that the difference of $\rsxi{\phi}-\rs{\phi}$, $\Axi{\phi}{x}{k}-\A{\phi}{x}{k}$, and $\Asxi{\phi}{x}{k}-\As{\phi}{x}{k}$ is lower bounded by $\zeta_{min}= \min_{i\in \{1,\hdots,N_\mu \}} \zeta_i$. \hfill\ensuremath{\blacksquare}
\end{pf}

Next, assume that $\boldsymbol{z}(k)$ is an affine function in $\boldsymbol{x}$ as in \eqref{eq:stacked_pred}, i.e., $\boldsymbol{z}(k) := C\boldsymbol{x}(k)+\boldsymbol{c}$. In Corollary \ref{cor1}, an increase in the predicate vector from $\boldsymbol{z}(k)$ to $\boldsymbol{z}_{\boldsymbol{\zeta}}(k)$ was induced by $\boldsymbol{\zeta}$. If an increase in $\boldsymbol{z}(k)$ is instead induced by a change of $\boldsymbol{x}(k)$ in $\boldsymbol{z}(k) := C\boldsymbol{x}(k)+\boldsymbol{c}$, then robustness can be interpreted in terms of the distance between hyperplanes. Note that for each predicate function $z_i(k):=C(i,:)\boldsymbol{x}(k)+c_i$ with $i\in \{1,\hdots,N_\mu \}$, we can define a corresponding time-dependent hyperplane $\mathcal{Z}_{i}(k):=\{\boldsymbol{x}\in\mathbb{R}^n|\boldsymbol{n}_i^T\boldsymbol{x}=e_i(k)\}$ with $\boldsymbol{n}_i^T:=C(i,:)$ and $e_i(k):=z_i(k)-c_i$. Now, consider two different signals $\boldsymbol{x}^\prime$ and $\boldsymbol{x}^{\prime\prime}$ resulting in two different versions $z_{i}^\prime(k):=C(i,:)\boldsymbol{x}^\prime(k)+c_i$ and $z_{i}^{\prime\prime}(k):=C(i,:)\boldsymbol{x}^{\prime\prime}(k)+c_i$ of the $i$-th predicate function. The  resulting two hyperplanes are $\mathcal{Z}_{i}^\prime(k):=\{\boldsymbol{x}\in\mathbb{R}^n|\boldsymbol{n}_i^T\boldsymbol{x}=e_{i}^\prime(k)\}$ and $\mathcal{Z}_{i}^{\prime\prime}(k):=\{\boldsymbol{x}\in\mathbb{R}^n|\boldsymbol{n}_i^T\boldsymbol{x}=e_{i}^{\prime\prime}(k)\}$ with $e_{i}^\prime(k):=z_{i}^\prime(k)-c_i$ and $e_{i}^{\prime\prime}(k):=z_{i}^{\prime\prime}(k)-c_i$. The distance between these two parallel hyperplanes at time $k$ is given by $\delta_{\mathcal{Z}_{i}^\prime,\mathcal{Z}_{i}^{\prime\prime}}(k):=\frac{|e_{i}^\prime(k)-e_{i}^{\prime\prime}(k)|}{\| \boldsymbol{n}_i \|}=\frac{|z_{i}^\prime(k)-z_{i}^{\prime\prime}(k)|}{\| \boldsymbol{n}_i \|}$. Assume again that $\phi$ contains the predicates $\mathcal{P}:=\{\mu_1,\hdots,\mu_{N_\mu}\}$ and consider the two predicate vectors $\boldsymbol{z}^\prime(k):=\begin{bmatrix} z_{1}^\prime(k) & z_{2}^\prime(k) & \hdots & z_{N_\mu}^\prime(k) \end{bmatrix}^T=C\boldsymbol{x}^\prime(k)+\boldsymbol{c}$ and $\boldsymbol{z}^{\prime\prime}(k):=\begin{bmatrix} z_{1}^{\prime\prime}(k) & z_{2}^{\prime\prime}(k) & \hdots & z_{N_\mu}^{\prime\prime}(k) \end{bmatrix}^T=C\boldsymbol{x}^{\prime\prime}(k)+\boldsymbol{c}$. Similarly to Corollary \ref{cor1}, define $\rsxie{\phi}$, $\Axie{\phi}{x}{k}$, and $\Asxie{\phi}{x}{k}$ if $\boldsymbol{z}^\prime(k)$ is used and $\rsxiz{\phi}$, $\Axiz{\phi}{x}{k}$, and $\Asxiz{\phi}{x}{k}$ if $\boldsymbol{z}^{\prime\prime}(k)$ is used in Definitions \ref{def:2}, \ref{def:44}, and \ref{def:55}. Next, Corollary \ref{cor2} provides a geometric interpretation since the robustness increase given by $\rsxiz{\phi}-\rsxie{\phi}$, $\Axiz{\phi}{x}{k}-\Axie{\phi}{x}{k}$, and $\Asxiz{\phi}{x}{k}-\Asxie{\phi}{x}{k}$ is proportional to the minimum distance of the hyperplanes  $\mathcal{Z}_{i}^\prime(k^*)$ and $\mathcal{Z}_{i}^{\prime\prime}(k^*)$ over all $i\in\{1,\hdots,N_\mu\}$ and $k\in\{k,\hdots,k+h_d^\phi\}$. Together, Corollary \ref{cor1} and \ref{cor2} state that robustness increases if $\boldsymbol{z}(k)$ increases element-wise. 
\begin{corollary}\label{cor2}  
Consider a formula $\phi$ in PNF and two signals $\boldsymbol{x}^\prime$ and $\boldsymbol{x}^{\prime\prime}$ such that $\boldsymbol{z}^\prime(k^*)<\boldsymbol{z}^{\prime\prime}(k^*)$ for all $k^*\in\{k,\hdots,k+h_d^\phi\}$. It then holds that $\rsxie{\phi}<\rsxiz{\phi}$, $\Axie{\phi}{x}{k}<\Axiz{\phi}{x}{k}$, and $\Asxie{\phi}{x}{k}<\Asxiz{\phi}{x}{k}$. Furthermore, it holds that $\rsxiz{\phi}-\rsxie{\phi}\ge \delta_{min}$, $\Axiz{\phi}{x}{k}-\Axie{\phi}{x}{k}\ge \delta_{min}$, and $\Asxiz{\phi}{x}{k}-\Asxie{\phi}{x}{k}\ge \delta_{min}$ where $\delta_{min}:=\min_{i\in \{1,\hdots,N_\mu\}}\min_{k^*\in\{k,\hdots,k+h_d^\phi\}}|z_{i}^{\prime}(k^*)-z_{i}^{\prime\prime}(k^*)|=\min_{i\in \{1,\hdots,N_\mu\}}\min_{k^*\in\{k,\hdots,k+h_d^\phi\}} \delta_{\mathcal{Z}_{i}^\prime,\mathcal{Z}_{i}^{\prime\prime}}(k^*)\|\boldsymbol{n}_i\|$. 
\end{corollary}
\begin{pf}
The proof is again based on the inductive definitions of SR, DASR, DSASR, and the fact that $\phi$ is in PNF. However, now we need to consider the minimum of $|z_{i}^\prime(k^*)-z_{i}^{\prime\prime}(k^*)|$ over all $i\in \{1,\hdots,N_\mu\}$ and $k^*\in \{k,\hdots,k+h_d^\phi\}$ to get the lower bound $\delta_{min}$.  \hfill\ensuremath{\blacksquare}
\end{pf}

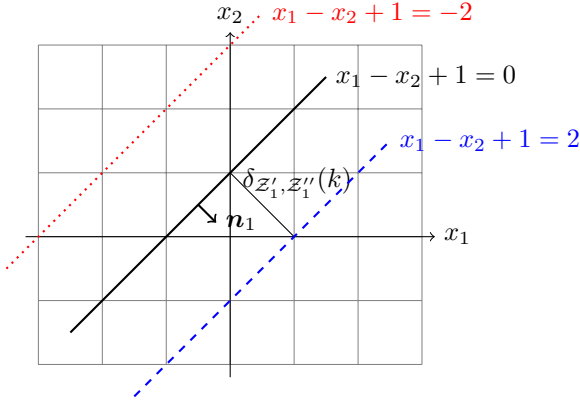
\begin{figure}[!t]
\centering
\begin{tikzpicture}[scale=0.85]
    \draw[very thin,color=gray] (-3,-2) grid (3,3);
    \draw[->,thin] (-3.2,0) -- (3.2,0) node[right] {$x_1$};
    \draw[->,thin] (0,-2.2) -- (0,3.2) node[above] {$x_2$};
    \coordinate (y) at (-2.5,-1.5);
    \coordinate (x) at (1.5,2.5);
    \coordinate (y1) at (-3.5,-0.5);
    \coordinate (x1) at (0.5,3.5);
    \coordinate (y2) at (-1.5,-2.5);
    \coordinate (x2) at (2.5,1.5);
    \draw[->,color=black,thick] (-0.5,0.5) -- (-0.22,0.22)node[right] {$\boldsymbol{n}_1$};
    \draw[-,color=black,thin] (0,1) -- (1,0);
    \node[] at (1.05,0.85) {$\delta_{\mathcal{Z}_{1}^\prime,\mathcal{Z}_{1}^{\prime\prime}}(k)$};
    \draw[-,color=black,thick] (y) to (x)node[right] {$x_1-x_2+1=0$};
    \draw[-,dotted,color=red,thick] (y1) to (x1)node[right] {$x_1-x_2+1=-2$};
    \draw[-,dashed,color=blue,thick] (y2) to (x2)node[right] {$x_1-x_2+1=2$};
\end{tikzpicture}
\caption{Consider $z_1(k):=x_1(k)-x_2(k)+1:=\boldsymbol{n}_1^T\boldsymbol{x}(k)+c_1$. The solid hyperplane separates the state space into two regions, one increasing and one decreasing robustness in the sense of Corollary \ref{cor2}. Moving the predicate $z_1(k)$ in $-\boldsymbol{n}_1$-direction from the solid to the dotted hyperplane, decreases robustness, while robustness is increased when moving into the direction of the dashed hyperplane.}
\label{fig:hyperplanes}
\end{figure}

\begin{example}\label{example_1}
Fig. \ref{fig:hyperplanes} displays three evaluations of the predicate function $z_1(k):=x_1(k)-x_2(k)+1:=\boldsymbol{n}_1^T\boldsymbol{x}(k)+c_1$, which is associated with the predicate $\mu_1$. Moving from $\boldsymbol{x}^\prime(k)\in \mathcal{Z}_{1}^\prime(k):=\{\boldsymbol{x}\in\mathbb{R}^n|\boldsymbol{n}_1^T\boldsymbol{x}=e_{1}^\prime(k)\}$ (solid hyperplane) with $e_{1}^\prime(k):=z_{1}^\prime(k)-c_1:=-1$ to $\boldsymbol{x}^{\prime\prime}(k)\in \mathcal{Z}_{1}^{\prime\prime}(k):=\{\boldsymbol{x}\in\mathbb{R}^n|\boldsymbol{n}_1^T\boldsymbol{x}=e_{1}^{\prime\prime}(k)\}$ (dashed hyperplane) with $e_{1}^{\prime\prime}(k):=z_{1}^{\prime\prime}(k)-c_1:=1$ results in an increase of the predicate function from  $z_{1}^{\prime}(k)=0$ to $z_{1}^{\prime\prime}(k)=2$. The distance $\delta_{\mathcal{Z}_{1}^\prime,\mathcal{Z}_{1}^{\prime\prime}}(k)$ between the solid and the dashed hyperplane is given by $\delta_{\mathcal{Z}_{1}^\prime,\mathcal{Z}_{1}^{\prime\prime}}(k)=\frac{|e_{1}^\prime(k)-e_{1}^{\prime\prime}(k)|}{\| \boldsymbol{n}_1 \| }=\sqrt{2}$ since $\| \boldsymbol{n}_1 \|=\sqrt{2}$. Next, consider the formula $\phi:=G_{[ a,b]}\mu_1$ and assume that $|z_{1}^\prime(k^*)-z_{1}^{\prime\prime}(k^*)|\ge 2$ for all $k^*\in\{k,\hdots,k+h_d^\phi\}$, which means that the hyperplane distance is $\delta_{\mathcal{Z}_{1}^\prime,\mathcal{Z}_{1}^{\prime\prime}}(k^*)\ge \sqrt{2}$ for all $k^*\in\{k,\hdots,k+h_d^\phi\}$.  According to Corollary  \ref{cor2}, it holds that $\rsxiz{\phi}-\rsxie{\phi}\ge \delta_{min}$, $\Axiz{\phi}{x}{k}-\Axie{\phi}{x}{k}\ge \delta_{min}$, and $\Asxiz{\phi}{x}{k}-\Asxie{\phi}{x}{k}\ge \delta_{min}$ with $\delta_{min}=2$. 
\end{example}

\subsection{Predicate Robustness Degree}
\label{sec:PRD}
 
Assume again nonlinear predicate functions $z_i(k):=f_i(\boldsymbol{x}(k))$ with $i\in\{1,\hdots,N_\mu\}$ as in \eqref{pred_nonlinear}. In the sequel, we will write $f_i(\boldsymbol{x}(k))$ to stress the dependence on $\boldsymbol{x}(k)$. Each predicate $\mu_i\in\mathcal{P}$ in $\phi$ is assigned a domain of influence at time $k$, denoted by $\mathcal{D}_{k,i}(\phi)$ and defined as 
\begin{align*}\mathcal{D}_{k,i}(\phi):=\mathcal{D}^\prime_{k,i}(\phi)\cup \mathcal{D}^\prime_{k,i}(\neg \phi)
\end{align*} 
with $\mathcal{D}^\prime_{k,i}(\phi):=\mathbb{N}\setminus\{k^\prime\in\mathbb{N}|\forall z\in\mathbb{R}, \forall\boldsymbol{x}^*\in f^{-1}_i(z), \exists \boldsymbol{x}\in\mathcal{L}_k(\phi) \text{ s.t. } \boldsymbol{x}(k^\prime)=\boldsymbol{x}^* \}$ where the set $f^{-1}_i(z):=\{\boldsymbol{x}\in\mathbb{R}^n|f_i(\boldsymbol{x})=z\}$ is the inverse image of $f_i(\boldsymbol{x})$. This domain of influence states for which $k^\prime\in\mathbb{N}$ the function values of $f_i(\boldsymbol{x}(k^\prime))$ can change the evaluation of $(\boldsymbol{x},k)\models \phi$. In other words, $f_i(\boldsymbol{x}(k^\prime))$ can take arbitrary values for each $k^\prime\notin \mathcal{D}_{k,i}(\phi)$ while the boolean evaluation of $(\boldsymbol{x},k)\models \phi$ will always be the same. For example, consider again the formula $\phi:=G_{[a,b]}\mu_1$ where $\mathcal{D}_{k,1}(\phi)=\kl \tau_k+a,\tau_k+b \kr$. In order to proceed, define the function $
h(x):=\begin{cases}
1 \text{ if } x\ge 0\\
0 \text{ if } x<0
\end{cases}$ and abbreviate $h_i^{+}(k^\prime):=h\big(f_i(\boldsymbol{x}(k^\prime))\big)$ and $h_i^{-}(k^\prime):=h\big(-f_i(\boldsymbol{x}(k^\prime))\big)$. 

\begin{definition}\label{def:prd}
Given a formula $\phi$ in PNF and a signal $\boldsymbol{x}$, the predicate robustness degree at time $k$ is defined as
\begin{align*}
\begin{split}
\mathcal{PRD}_k(\boldsymbol{x},\phi)&:=\\
&\hspace{-1.5cm}\begin{cases}
\sum\limits_{i=1}^{N_\mu}\sum\limits_{k^\prime\in \mathcal{D}_{k,i}(\phi)} {h_i^-}(k^\prime)f_i(\boldsymbol{x}(k^\prime)) \text{ if } \boldsymbol{x}\notin \mathcal{L}_k(\phi)\\
\sum\limits_{i=1}^{N_\mu}\sum\limits_{k^\prime\in \mathcal{D}_{k,i}(\phi)} h_i^+(k^\prime)f_i(\boldsymbol{x}(k^\prime)) \text{ if } \boldsymbol{x}\in \mathcal{L}_k(\phi)
\end{cases}
\end{split}
\end{align*}
\end{definition}
\vspace{-0.5cm}
The predicate robustness degree measures robustness by considering predicates as follows: if $\boldsymbol{x}\in \mathcal{L}_k(\phi)$, we sum over all $f_i(\boldsymbol{x}(k^\prime))> 0$ with $i\in\{1,\hdots,N_\mu\}$ and $k^\prime\in \mathcal{D}_{k,i}(\phi)$. If $\boldsymbol{x}\notin \mathcal{L}_k(\phi)$, we instead sum over all $f_i(\boldsymbol{x}(k^\prime))< 0$ with $i\in\{1,\hdots,N_\mu\}$ and $k^\prime\in \mathcal{D}_{k,i}(\phi)$. Note that if $f_i(\boldsymbol{x}(k^\prime))\ge 0$ for all $i\in \{1,\hdots,N_\mu\}$ and $k^\prime\in\mathcal{D}_{k,i}(\phi)$, then $(\boldsymbol{x},k)\models\phi$ holds. Conversely, if $f_i(\boldsymbol{x}(k^\prime))< 0$ for all $i\in \{1,\hdots,N_\mu\}$ and $k^\prime\in\mathcal{D}_{k,i}(\phi)$, then $(\boldsymbol{x},k)\not\models\phi$, i.e., $(\boldsymbol{x},k)\models\neg\phi$, holds. Hence, the predicate robustness degree $\mathcal{PRD}_k(\boldsymbol{x},\phi)$ measures the summed distance of all $f_i(\boldsymbol{x}(k^\prime))>0$ (if $\boldsymbol{x}\in \mathcal{L}_k(\phi)$) or $f_i(\boldsymbol{x}(k^\prime))<0$ (if $\boldsymbol{x}\notin \mathcal{L}_k(\phi)$) with $i\in \{1,\hdots,N_\mu\}$ and $k^\prime\in\mathcal{D}_{k,i}(\phi)$ to $f_i(\boldsymbol{x}(k^\prime))=0$; $f_i(\boldsymbol{x}(k^\prime))=0$ is the value of the predicate where the evaluation of $(\boldsymbol{x},k)\models\phi$ will change its truth value if all $f_i(\boldsymbol{x}(k))>0$ (if $\boldsymbol{x}\in \mathcal{L}_k(\phi)$) or $f_i(\boldsymbol{x}(k))<0$ (if $\boldsymbol{x}\notin \mathcal{L}_k(\phi)$) are changed such that $f_i(\boldsymbol{x}(k^\prime))=0$.

\begin{remark}\label{remark4}
The predicate robustness degree is suitable for comparing two signals $\boldsymbol{x}^\prime$ and $\boldsymbol{x}^{\prime\prime}$ with respect to a formula $\phi$. It is, however, not suited for considerations of the worst case disturbance (robustness degree).
\end{remark}

\begin{example}\label{example2}
Consider the discrete-time dynamics
\begin{align*}
\boldsymbol{x}(k+1) = \begin{bmatrix}
0.79 & 0\\
0.176 & 0.0296
\end{bmatrix}\boldsymbol{x}(k)+
\begin{bmatrix}
0.281\\
0.0296
\end{bmatrix}u(k)
\end{align*}
that represent a coupled two-tank process  with $T:=12$. The input $u$ represents a pump that fills the first tank with water (water level indicated by $x_1$), while there is an outlet in the first tank from which water pours into the second tank (water level indicated by $x_2$). There is another outlet in the second tank so that water seeps into the ground. We impose the specification $\phi:=G_{[ 144,216 ]}(x_1\ge 1) \wedge G_{[ 372,444 ]}(x_1\ge 1)$. It holds that $\A{\phi}{x}{k}=\As{\phi}{x}{k}$ since the semantics for the always-operator are the same for DASR and DSASR, i.e., the notions of DASR and DSASR coincide in this example. We solve two optimization problems to demonstrate the difference between the use of $\rss{\phi}{k}$ (SR) and $\A{\phi}{x}{k}$ (DASR):  1) $\underset{\boldsymbol{u}_{st}}{\operatorname{argmax}} \; \rss{\phi}{0}$ and 2) $\underset{\boldsymbol{u}_{st}}{\operatorname{argmax}} \; \A{\phi}{x}{0}$ s.t. $x_1(k^\prime)\ge 1$ for all $k^\prime \in \kl  144,216 \kr\cup\kl 372,444 \kr$. Additionally, both 1) and 2) are subject to the constraints $0\le u(k^\prime) \le 3$ for all $k^\prime \in \kl 0 , 444 \kr$ and $\sum_{k^\prime=0}^{h_d^\phi} u(k^\prime)\le 20$. The result is shown in Fig. \ref{fig:dasr_sr} and it is visible that DASR gives a greater average satisfaction within the regions $[144,216]$ and $[372,444]$ resulting in a greater predicate robustness degree than SR. Note that the red shaded areas show where a signal could potentially violate the formula $\phi$. The predicate robustness degree at $k:=0$ is $\mathcal{PRD}_0(\boldsymbol{x},\phi)=5.12$ for DASR and $\mathcal{PRD}_0(\boldsymbol{x},\phi)=3.01$ for SR, whereas the robustness degree at $k:=0$ is $\mathcal{RD}_0(\boldsymbol{x},\phi)=0$ for DASR and $\mathcal{RD}_0(\boldsymbol{x},\phi)=0.215$ for SR. Hence, a small disturbance at exactly $216$ seconds (see Fig. \ref{fig:dasr_sr}), i.e., a worst case disturbance for both shown signals,  will not change $(\boldsymbol{x},0)\models \phi$ for the SR approach, but it can for the DASR approach. On the other hand, if there is a disturbance between $150$ and $200$ seconds, the DASR approach can tolerate a bigger disturbance than the SR approach.
\begin{figure}
\centering
%
%
\definecolor{mycolor1}{rgb}{0.00000,0.44700,0.74100}%
\definecolor{mycolor2}{rgb}{0.85000,0.32500,0.09800}%
\begin{tikzpicture}[scale=0.33]
\begin{axis}[%
width=6.028in,
height=4.754in,
at={(1.011in,0.642in)},
scale only axis,
xmin=0,
xmax=444,
xlabel={Time (s)},
ylabel={$x_1$},
xmajorgrids,
ymin=-0.05,
ymax=1.8,
ymajorgrids,
axis background/.style={fill=white},
legend style={legend cell align=left,align=left,draw=white!15!black},
title style={font=\LARGE},xlabel style={font=\Huge},ylabel style={font=\Huge},legend style={font=\LARGE},ticklabel style={font=\LARGE},
]
\addplot [color=black,solid,line width=2.0pt]
  table[row sep=crcr]{%
0	0\\
12	2.3762783563086e-10\\
24	4.32518119494922e-10\\
36	5.96165920512152e-10\\
48	7.38813665387434e-10\\
60	8.70491377573036e-10\\
72	1.00261818705011e-09\\
84	1.15122347267928e-09\\
96	1.34538065676076e-09\\
108	1.65756962385154e-09\\
120	2.39720837361701e-09\\
132	0.198734178558635\\
144	1.00000000048831\\
156	1.50854170425861\\
168	1.73298867056923\\
180	1.65270606961076\\
192	1.44476490876355\\
204	1.22163877784645\\
216	1.00000000018025\\
228	0.790000000374642\\
240	0.624100000533593\\
252	0.493039000666329\\
264	0.389500810780876\\
276	0.307705640784736\\
288	0.243087456506779\\
300	0.192039090955314\\
312	0.15171088221394\\
324	0.119851597385223\\
336	0.094682762530346\\
348	0.0747993834975956\\
360	0.198734178564345\\
372	1.00000000046865\\
384	1.50798887944502\\
396	1.7333664475318\\
408	1.65368467774722\\
420	1.44471455045453\\
432	1.22162030319116\\
444	1.0000000001702\\
};
\addlegendentry{DASR};

\addplot [color=blue,dashed,line width=2.0pt]
  table[row sep=crcr]{%
0	0\\
12	1.4966811470174e-09\\
24	2.72334082666005e-09\\
36	3.75218860699459e-09\\
48	4.64735903708244e-09\\
60	5.4713039748485e-09\\
72	6.29451499689054e-09\\
84	7.21488979068373e-09\\
96	8.40813948996469e-09\\
108	1.030822517918e-08\\
120	1.47218957101859e-08\\
132	0.471449954089412\\
144	1.21544545912359\\
156	1.21544546158293\\
168	1.21544546153008\\
180	1.21544546158957\\
192	1.21544546157585\\
204	1.2154454614356\\
216	1.21544545804098\\
228	0.96020191332018\\
240	0.758559513025436\\
252	0.599262016838789\\
264	0.473416994913823\\
276	0.373999427679383\\
288	0.295459549686777\\
300	0.233413046254038\\
312	0.184396308827172\\
324	0.145673086751663\\
336	0.115081742312267\\
348	0.0909145833029123\\
360	0.471449959290729\\
372	1.21544545920551\\
384	1.21544546173545\\
396	1.21544546154537\\
408	1.2154454615936\\
420	1.21544546158561\\
432	1.21544546142782\\
444	1.21544545797816\\
};
\addlegendentry{SR};

\draw[draw=none,fill=red!30]  (372,0) -- (372,105) -- (444,105) -- (444,0) -- (372,0);
\draw[draw=none,fill=red!30]  (144,0) -- (144,105) -- (216,105) -- (216,0) -- (144,0);
\end{axis}
\end{tikzpicture}%
\caption{For the water level $x_1$ of the first tank, we impose the specification $\phi:=G_{[ 144,216 ]}(x_1\ge 1) \wedge G_{[ 372,444 ]}(x_1\ge 1)$. SR performs better in terms of the worst case disturbance, while DASR can, on average, tolerate more disturbance.}
\label{fig:dasr_sr}
\end{figure}
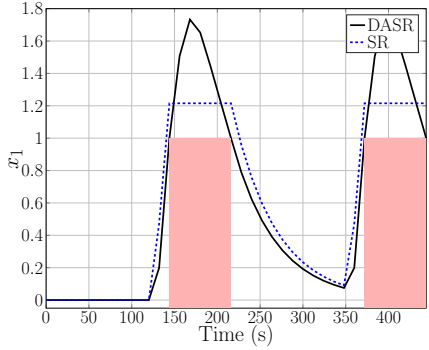  
\end{example}

One may now wonder when to choose which notion. Putting computational issues aside for a moment, it may be more suitable to use SR for safety-critical systems. For systems that are not safety-critical or where constraints are mostly soft, the notion of DASR/DSASR may be a good alternative.  A worst case disturbance may now more easily lead to not satisfying the formula; however, for other disturbances the system may perform more robustly as illustrated in Example~\ref{example2}. Note that maximizing DASR/DSASR leads to increasing the predicate robustness degree. DASR may lead to a higher predicate robustness degree than SR. This intuitive observation can be justified by looking at the definitions of the always- and until-operator for both DASR and SR. While maximizing SR leads to maximizing the worst case, i.e., only one point in time (min-operation for until- and always-operators in Definition~\ref{def:2}), DASR maximizes predicates over certain time intervals (summation-operator for until- and always-operators in Definition~\ref{def:44}). These time intervals are linked to $\mathcal{D}_{k,i}(\phi)$, which are used to calculate $\mathcal{PRD}_k(\boldsymbol{x},\phi)$.

\section{Problem Statement}
\label{sec:problem_statement}

In this paper, the STL fragment under consideration is 
\begin{subequations}\label{class:all}
\begin{align}
\gamma&::=  \mu \;|\; \neg \mu \;|\;  \gamma_1 \wedge \gamma_2 \label{class:mu} \\
\psi&::= \gamma_1 \until{a}{b} \gamma_2 \;|\; F_{[ a,b]} \gamma_1 \;|\; G_{[ a,b]}\gamma_1\;|\; \psi_1 \wedge \psi_2 \;|\; \psi_1 \vee \psi_2\label{class:psi}\\
\phi &::= G_{[0,\infty]}\psi \;|\; event  \implies \psi. \label{class:phi}
\end{align}
\end{subequations} 
where $\gamma_1$ and $\gamma_2 $ are formulas of class $\gamma$ given in \eqref{class:mu}, whereas $\psi_1$ and $\psi_2$ are formulas of class $\psi$ given in \eqref{class:psi}. The formulas $\phi:=G_{[ 0,\infty ]}\psi$ are called \emph{all-time satisfying formulas}, which means that the formula $\psi$ is imposed at every sampling step; $\phi:=event  \implies \psi$ are called \emph{one-time satisfying formulas} since $\psi$ is to be satisfied once. The boolean variable $event\in\{\top,\bot\}$ is an indicator for the time $k_{event}$ when $\psi$ is activated, i.e., $event(k_{event})=\top$ and $event(k)=\bot$ if $k\in\mathbb{N}\setminus k_{event}$. This means that $(\boldsymbol{x},0)\models\phi$ holds if and only if $(\boldsymbol{x},k_{event})\models \psi$. We argue that the STL fragment in \eqref{class:phi} offers a good trade-off between computational tractability and expressivity and allows to express many real-world, especially robotic, specifications. Periodic tasks, such as surveillance, can be formulated as all-time satisfying formulas, while one-time satisfying formulas can be used to express reachability or other robotic tasks such as formation control. The GR(1) fragment of LTL introduced in \cite{piterman2006synthesis} also offers computationally tractable algorithms, while considering a reactive design procedure. Compared to the GR(1) fragment, we allow for quantitative temporal requirements, i.e., hard deadlines can be imposed, while also considering until- and disjunction-operators.

Define next two variables $k_l, k_h\in\mathbb{N}$. For \emph{all-time satisfying formulas}, set $k_l:=k_0-h_d^{\psi}+1$ and $k_h:=k_0+N-h_d^{\psi}$ where $N$ is a prediction horizon, $k_0$ is the current time, and $h_d^{\psi}$ is the discrete formula length of $\psi$. Knowledge of $\boldsymbol{x}(k)$ for all $k\in\{k_l,k_l+1,\hdots,k_0+N\}$ is sufficient to evaluate $(\boldsymbol{x},k^\prime)\models \psi$ for all $k^\prime\in \{k_l,k_l+1,\hdots,k_h\}$. This means that satisfaction of $\psi$ is considered for the past and the future with respect to $k_0$, which resembles the notion of past satisfaction in \cite{sadraddini,raman2}. For \emph{one-time satisfying formulas}, set $k_l:=k_h:=k_{event}$ since only $(\boldsymbol{x},k_{event})\models \psi$ is required. We assume that the time intervals $[a,b]$ are finite and that $N\ge h_d^\psi$. To obtain computationally efficient algorithms to control the system \eqref{system_discrete} subject to an STL formula $\phi$ with a corresponding formula $\psi$ as in \eqref{class:all}, neither $\rs{\psi}$ nor $\A{\psi}{x}{k}$ are used and instead $\As{\psi}{x}{k}$ is used. Note that the $k_1$'s in Definition~\ref{def:55} are needed in order to fully define $\As{\psi}{x}{k}$. Therefore, we first propose an offline algorithm, referred to as Algorithm 1, that calculates candidate $k_1$ for each temporal operator in $\psi$ that then define the DSASR semantics $\As{\psi}{x}{k}$. In a second step, $\As{\psi}{x}{k}$ is maximized through an optimization problem, while the obtained candidate $k_1$'s are additionally used to ensure that $(\boldsymbol{x},k^\prime)\models \psi$ for all  $k^\prime\in\{k_l,k_l+1,\hdots,k_h\}$ as explained more in Section \ref{sec:our_approach}.

\begin{problem}\label{problem1}
Given an STL formula $\phi$ as in \eqref{class:phi}, the corresponding DSASR semantics $\As{\psi}{x}{k}$ that are to be obtained by Algorithm 1, the system  \eqref{system_discrete}, the current state $\boldsymbol{x}(k_0)$, and a prediction horizon $N\ge h^{\psi}_d$, compute
\begin{subequations}\label{eq:problem1}
\begin{align}
&\underset{\boldsymbol{u}_{st}}{\operatorname{argmax}} \sum_{k^\prime=k_l}^{k_h} \As{\psi}{x}{k^\prime} -\sum_{k^\prime=k_0}^{k_0+N-1}\|\boldsymbol{u}(k^\prime) \|^2_M \label{eq:cost_111}\\
\text{s.t. } &\boldsymbol{x}(k+1) = A\boldsymbol{x}(k)+B\boldsymbol{u}(k) \label{eq:constraint_x}\\
&A_{\boldsymbol{u}} \boldsymbol{u}(k^\prime)\le \boldsymbol{b}_{\boldsymbol{u}} \;\;\; \forall k^\prime\in\{k_0,\hdots,k_0+N-1\}\label{eq:input_constraint}\\
&A_{\boldsymbol{x}} \boldsymbol{x}(k^\prime)\le \boldsymbol{b}_{\boldsymbol{x}} \;\;\; \forall k^\prime\in\{k_0+1,\hdots,k_0+N\}\label{eq:state_constraint}\\
&(\boldsymbol{x},k^\prime) \models \psi \;\;\forall k^\prime\in\kl \tau_{k_l},\tau_{k_h} \kr,\label{eq:constraint_problem}
\end{align}
\end{subequations}
where $A_{\boldsymbol{u}}$, $A_{\boldsymbol{x}}$, $\boldsymbol{b}_{\boldsymbol{u}}$, and $\boldsymbol{b}_{\boldsymbol{x}}$ are of appropriate size and may be used to define input and state constraints, while $M\in\mathbb{R}^{mn\times mn}$ is a positive semidefinite matrix.
\label{prob:gen}
\end{problem}
In the remainder, \eqref{eq:constraint_x}, \eqref{eq:input_constraint}, \eqref{eq:state_constraint}, and the term $\sum_{k^\prime=k_0}^{k_0+N-1}\|{\boldsymbol{u}(k^\prime)}\|^2_M$ in \eqref{eq:cost_111} are not explicitly mentioned. Solving \eqref{eq:problem1} in a receding horizon fashion will result in a closed-loop solution. The procedure is initialized at time $k_0:=0$ where \eqref{eq:problem1} is solved. Only the first element of $\boldsymbol{u}_{st}$ is applied to the system and the procedure is repeated at the next sampling step with $k_0$ increased by $1$.
\begin{corollary}\label{cor3}
The closed-loop solution obtained by iteratively solving Problem \ref{problem1} results in the satisfaction of $\phi$, i.e., $(\boldsymbol{x},0)\models \phi$,  if \eqref{eq:problem1} is feasible at every sampling step.
\end{corollary}

\begin{pf}
The assumption that \eqref{eq:problem1} is feasible at every sampling step is necessary. For \emph{one-time satisfying formulas}, the result trivially holds. For \emph{all-time satisfying formulas}, it holds that $(\boldsymbol{x},0)\models \phi$ if and only if $(\boldsymbol{x},k^\prime)\models \psi$ for all $k^\prime\in \mathbb{N}$. Due to the choice of $k_l$, we consider past satisfaction and the result follows.  A similar idea is used in \cite{raman2} where the past $h_d^{\psi}-1$ inputs and robustness constraints are stored and considered. \hfill\ensuremath{\blacksquare}
\end{pf}

\section{Control Strategy}
\label{sec:control_strategy}

Section \ref{sec:k1} describes the offline calculation of $k_1$ to obtain the DSASR semantics, while Section \ref{sec:our_approach} explains our proposed control strategy that depends on these $k_1$. 

\subsection{Offline calculation of $k_1$}
\label{sec:k1}

The parameter $k_1$ needs to be calculated for each eventually- and until-operator in $\phi$ (see Definition \ref{def:55}) and hence $k_1^i$ needs to be calculated for each $i\in \{1,\dots,N^\phi\}$, where $N^\phi$ is the total number of eventually- and until-operators contained in $\phi$; $k_1^i$ is therefore the $k_1$ of the $i$-th eventually- and until-operator. We first provide two examples that give some intuition on this before we present Algorithm \ref{alg:1} that automatically calculates $k_1^i$. Since Algorithm \ref{alg:1} uses the discrete-time intervals $\kl a,b\kr$, our examples will directly use the discrete-time intervals, e.g., $F_{\{5,\hdots,15\}}\gamma_1$ instead of $F_{[5,15]}\gamma_1$ when $\kl 5,15 \kr=\{5,6\hdots,15\}$. For one-time satisfying formulas, $k_1^i$ is calculated as in the next example.

\begin{example}
Assume $\phi_1:=event \implies \psi_1$ with $\psi_1:=F_{\{5,\hdots,15\}}\gamma_1$ and $N^{\phi_1}=1$. Choosing $k_1^1\in\{k_{event}+5,\hdots,k_{event}+15\}$ results in $(\boldsymbol{x},0)\models\phi_1$, which is the same as $(\boldsymbol{x},k_{event})\models\psi_1$, if $(\boldsymbol{x},k_1^1)\models\gamma_1$ holds. If the number of eventually- and until-operators exceeds one, e.g., $N^{\phi}>1$, the selection of $k_1^i$ can be extended as follows: assume $\phi_2 := event \implies (\psi_1 \wedge \psi_2)$ with $\psi_1$ as before and $\psi_2:=F_{\{5,\hdots,15\}}\gamma_2$. Consequently, $N^{\phi_2}=2$ and $k_1^i$ needs to be selected for $i\in \{1,2\}$. In this case, $k_1^{1}\in\{k_{event}+5,\hdots,k_{event}+10\}$ and $k_1^{2}\in\{k_{event}+11,\hdots,k_{event}+15\}$ can be used for the first and second subformula, respectively. If now $(\boldsymbol{x},k_1^1)\models\gamma_1$ and $(\boldsymbol{x},k_1^2)\models\gamma_2$ hold, then $(\boldsymbol{x},k_{event})\models\psi_1\wedge\psi_2$ holds and consequently $(\boldsymbol{x},0)\models\phi_2$ is true. The reason why $k_1^1\neq k_1^2$, becomes obvious when $\gamma_1:=x \ge 2$ and $\gamma_2:=x\le 1$. 
\end{example}

\begin{figure*}
\centering
\begin{subfigure}{0.38\textwidth}\caption{Calculation of $k_1^1(k^\prime)$ for $\phi_1=G_{[0,\infty]}\psi_1$}\label{fig:t1_precalc_idea}
%
%

\begin{tikzpicture}[scale=0.155]
\pgfmathsetmacro{\y}{0.75}
\pgfmathsetmacro{\x}{1.5}
    \draw[->,thin] (0,0) -- (\x*0,\y*37) node[right] {$K_1^1(k^\prime)$};
    \draw[->,thin] (0,0) -- (\x*21,\y*0) node[above] {$k^\prime$};
	\draw [thick,color=black] plot [smooth] coordinates {(\x*0,\y*5)(\x*0,\y*15)};
	\draw [thick,color=black] plot [smooth] coordinates {(\x*1,\y*6)(\x*1,\y*16)};
	\draw [thick,color=black] plot [smooth] coordinates {(\x*2,\y*7)(\x*2,\y*17)};
	\draw [thick,color=black] plot [smooth] coordinates {(\x*3,\y*8)(\x*3,\y*18)};
	\draw [thick,color=black] plot [smooth] coordinates {(\x*4,\y*9)(\x*4,\y*19)};
	\draw [thick,color=black] plot [smooth] coordinates {(\x*5,\y*10)(\x*5,\y*20)};
	\draw [thick,color=black] plot [smooth] coordinates {(\x*6,\y*11)(\x*6,\y*21)};
	\draw [thick,color=black] plot [smooth] coordinates {(\x*7,\y*12)(\x*7,\y*22)};
	\draw [thick,color=black] plot [smooth] coordinates {(\x*8,\y*13)(\x*8,\y*23)};
	\draw [thick,color=black] plot [smooth] coordinates {(\x*9,\y*14)(\x*9,\y*24)};
	\draw [thick,color=black] plot [smooth] coordinates {(\x*10,\y*15)(\x*10,\y*25)};
	\draw [thick,color=black] plot [smooth] coordinates {(\x*11,\y*16)(\x*11,\y*26)};
	\draw [thick,color=black] plot [smooth] coordinates {(\x*12,\y*17)(\x*12,\y*27)};
	\draw [thick,color=black] plot [smooth] coordinates {(\x*13,\y*18)(\x*13,\y*28)};
	\draw [thick,color=black] plot [smooth] coordinates {(\x*14,\y*19)(\x*14,\y*29)};
	\draw [thick,color=black] plot [smooth] coordinates {(\x*15,\y*20)(\x*15,\y*30)};
	\draw [thick,color=black] plot [smooth] coordinates {(\x*16,\y*21)(\x*16,\y*31)};
	\draw [thick,color=black] plot [smooth] coordinates {(\x*17,\y*22)(\x*17,\y*32)};
	\draw [thick,color=black] plot [smooth] coordinates {(\x*18,\y*23)(\x*18,\y*33)};
	\draw [thick,color=black] plot [smooth] coordinates {(\x*19,\y*24)(\x*19,\y*34)};
	\draw [thick,color=black] plot [smooth] coordinates {(\x*20,\y*25)(\x*20,\y*35)};
	\node at (\x*-0.95,\y*0) {$0$};
	\node at (\x*-0.95,\y*5) {$5$};
	\node at (\x*-0.95,\y*10) {$10$};
	\node at (\x*-0.95,\y*15) {$15$};
	\node at (\x*-0.95,\y*20) {$20$};
	\node at (\x*-0.95,\y*25) {$25$};
	\node at (\x*-0.95,\y*30) {$30$};
	\node at (\x*-0.95,\y*35) {$35$};
	\node at (\x*0,\y*-2) {$0$};
	\node at (\x*5,\y*-2) {$5$};
	\node at (\x*10,\y*-2) {$10$};
	\node at (\x*15,\y*-2) {$15$};
	\node at (\x*20,\y*-2) {$20$};
	\draw [very thin,color=gray] plot [smooth] coordinates {(\x*0,\y*0)(\x*20,\y*0)};
	\draw [very thin,color=gray] plot [smooth] coordinates {(\x*0,\y*5)(\x*20,\y*5)};
	\draw [very thin,color=gray] plot [smooth] coordinates {(\x*0,\y*10)(\x*20,\y*10)};
	\draw [very thin,color=gray] plot [smooth] coordinates {(\x*0,\y*15)(\x*20,\y*15)};
	\draw [very thin,color=gray] plot [smooth] coordinates {(\x*0,\y*20)(\x*20,\y*20)};
	\draw [very thin,color=gray] plot [smooth] coordinates {(\x*0,\y*25)(\x*20,\y*25)};
	\draw [very thin,color=gray] plot [smooth] coordinates {(\x*0,\y*30)(\x*20,\y*30)};
	\draw [very thin,color=gray] plot [smooth] coordinates {(\x*0,\y*35)(\x*20,\y*35)};
	\draw [very thin,color=gray] plot [smooth] coordinates {(\x*0,\y*0)(\x*0,\y*35)};
	\draw [very thin,color=gray] plot [smooth] coordinates {(\x*5,\y*0)(\x*5,\y*35)};
	\draw [very thin,color=gray] plot [smooth] coordinates {(\x*10,\y*0)(\x*10,\y*35)};
	\draw [very thin,color=gray] plot [smooth] coordinates {(\x*15,\y*0)(\x*15,\y*35)};
	\draw [very thin,color=gray] plot [smooth] coordinates {(\x*20,\y*0)(\x*20,\y*35)};
	\draw [postaction={draw,decorate,decoration={shape backgrounds,shape=circle,shape size=1.5mm,%
        shape sep={0.275cm, between center}},fill=black}] plot [thick,smooth] coordinates {(\x*0,\y*15)(\x*10,\y*15)};
        	\draw [postaction={draw,decorate,decoration={shape backgrounds,shape=circle,shape size=1.5mm,%
        shape sep={0.275cm, between center}},fill=black}] plot [thick,smooth] coordinates {(\x*11,\y*26)(\x*20,\y*26)};
	
\end{tikzpicture}
\end{subfigure}
\begin{subfigure}{0.53\textwidth}\caption{Calculation of $k_1^1(k^\prime)$ and $k_1^2(k^\prime)$ for $\phi_2=G_{[0,\infty]}(\psi_1 \wedge \psi_2)$ }\label{fig:t1_precalc_idea1}
%
%

\begin{tikzpicture}[scale=0.155]
\pgfmathsetmacro{\y}{0.75}
\pgfmathsetmacro{\x}{1}
    \draw[->,thin] (0,0) -- (\x*0,\y*37) node[right] {$K_1^1(k^\prime)$};
    \draw[->,thin] (\x*25,0) -- (\x*25,\y*37) node[right] {$K_1^2(k^\prime)$};
    \draw[->,thin] (0,0) -- (\x*21,\y*0) node[above] {$k^\prime$};
    \draw[->,thin] (\x*25,0) -- (\x*46,\y*0) node[above] {$k^\prime$};
	\draw [thick,color=black] plot [smooth] coordinates {(\x*0,\y*5)(\x*0,\y*15)};
	\draw [thick,color=black] plot [smooth] coordinates {(\x*1,\y*6)(\x*1,\y*16)};
	\draw [thick,color=black] plot [smooth] coordinates {(\x*2,\y*7)(\x*2,\y*17)};
	\draw [thick,color=black] plot [smooth] coordinates {(\x*3,\y*8)(\x*3,\y*18)};
	\draw [thick,color=black] plot [smooth] coordinates {(\x*4,\y*9)(\x*4,\y*19)};
	\draw [thick,color=black] plot [smooth] coordinates {(\x*5,\y*10)(\x*5,\y*20)};
	\draw [thick,color=black] plot [smooth] coordinates {(\x*6,\y*11)(\x*6,\y*21)};
	\draw [thick,color=black] plot [smooth] coordinates {(\x*7,\y*12)(\x*7,\y*22)};
	\draw [thick,color=black] plot [smooth] coordinates {(\x*8,\y*13)(\x*8,\y*23)};
	\draw [thick,color=black] plot [smooth] coordinates {(\x*9,\y*14)(\x*9,\y*24)};
	\draw [thick,color=black] plot [smooth] coordinates {(\x*10,\y*15)(\x*10,\y*25)};
	\draw [thick,color=black] plot [smooth] coordinates {(\x*11,\y*16)(\x*11,\y*26)};
	\draw [thick,color=black] plot [smooth] coordinates {(\x*12,\y*17)(\x*12,\y*27)};
	\draw [thick,color=black] plot [smooth] coordinates {(\x*13,\y*18)(\x*13,\y*28)};
	\draw [thick,color=black] plot [smooth] coordinates {(\x*14,\y*19)(\x*14,\y*29)};
	\draw [thick,color=black] plot [smooth] coordinates {(\x*15,\y*20)(\x*15,\y*30)};
	\draw [thick,color=black] plot [smooth] coordinates {(\x*16,\y*21)(\x*16,\y*31)};
	\draw [thick,color=black] plot [smooth] coordinates {(\x*17,\y*22)(\x*17,\y*32)};
	\draw [thick,color=black] plot [smooth] coordinates {(\x*18,\y*23)(\x*18,\y*33)};
	\draw [thick,color=black] plot [smooth] coordinates {(\x*19,\y*24)(\x*19,\y*34)};
	\draw [thick,color=black] plot [smooth] coordinates {(\x*20,\y*25)(\x*20,\y*35)};
	\draw [thick,color=black] plot [smooth] coordinates {(\x*25,\y*5)(\x*25,\y*15)};
	\draw [thick,color=black] plot [smooth] coordinates {(\x*26,\y*6)(\x*26,\y*16)};
	\draw [thick,color=black] plot [smooth] coordinates {(\x*27,\y*7)(\x*27,\y*17)};
	\draw [thick,color=black] plot [smooth] coordinates {(\x*28,\y*8)(\x*28,\y*18)};
	\draw [thick,color=black] plot [smooth] coordinates {(\x*29,\y*9)(\x*29,\y*19)};
	\draw [thick,color=black] plot [smooth] coordinates {(\x*30,\y*10)(\x*30,\y*20)};
	\draw [thick,color=black] plot [smooth] coordinates {(\x*31,\y*11)(\x*31,\y*21)};
	\draw [thick,color=black] plot [smooth] coordinates {(\x*32,\y*12)(\x*32,\y*22)};
	\draw [thick,color=black] plot [smooth] coordinates {(\x*33,\y*13)(\x*33,\y*23)};
	\draw [thick,color=black] plot [smooth] coordinates {(\x*34,\y*14)(\x*34,\y*24)};
	\draw [thick,color=black] plot [smooth] coordinates {(\x*35,\y*15)(\x*35,\y*25)};
	\draw [thick,color=black] plot [smooth] coordinates {(\x*36,\y*16)(\x*36,\y*26)};
	\draw [thick,color=black] plot [smooth] coordinates {(\x*37,\y*17)(\x*37,\y*27)};
	\draw [thick,color=black] plot [smooth] coordinates {(\x*38,\y*18)(\x*38,\y*28)};
	\draw [thick,color=black] plot [smooth] coordinates {(\x*39,\y*19)(\x*39,\y*29)};
	\draw [thick,color=black] plot [smooth] coordinates {(\x*40,\y*20)(\x*40,\y*30)};
	\draw [thick,color=black] plot [smooth] coordinates {(\x*41,\y*21)(\x*41,\y*31)};
	\draw [thick,color=black] plot [smooth] coordinates {(\x*42,\y*22)(\x*42,\y*32)};
	\draw [thick,color=black] plot [smooth] coordinates {(\x*43,\y*23)(\x*43,\y*33)};
	\draw [thick,color=black] plot [smooth] coordinates {(\x*44,\y*24)(\x*44,\y*34)};
	\draw [thick,color=black] plot [smooth] coordinates {(\x*45,\y*25)(\x*45,\y*35)};
	\node at (\x*-0.95,\y*0) {$0$};
	\node at (\x*-0.95,\y*5) {$5$};
	\node at (\x*-0.95,\y*10) {$10$};
	\node at (\x*-0.95,\y*15) {$15$};
	\node at (\x*-0.95,\y*20) {$20$};
	\node at (\x*-0.95,\y*25) {$25$};
	\node at (\x*-0.95,\y*30) {$30$};
	\node at (\x*-0.95,\y*35) {$35$};
	\node at (\x*0,\y*-2) {$0$};
	\node at (\x*5,\y*-2) {$5$};
	\node at (\x*10,\y*-2) {$10$};
	\node at (\x*15,\y*-2) {$15$};
	\node at (\x*20,\y*-2) {$20$};
	\node at (\x*24,\y*0) {$0$};
	\node at (\x*24,\y*5) {$5$};
	\node at (\x*24,\y*10) {$10$};
	\node at (\x*24,\y*15) {$15$};
	\node at (\x*24,\y*20) {$20$};
	\node at (\x*24,\y*25) {$25$};
	\node at (\x*24,\y*30) {$30$};
	\node at (\x*24,\y*35) {$35$};
	\node at (\x*25,\y*-2) {$0$};
	\node at (\x*30,\y*-2) {$5$};
	\node at (\x*35,\y*-2) {$10$};
	\node at (\x*40,\y*-2) {$15$};
	\node at (\x*45,\y*-2) {$20$};
	\draw [very thin,color=gray] plot [smooth] coordinates {(\x*0,\y*0)(\x*20,\y*0)};
	\draw [very thin,color=gray] plot [smooth] coordinates {(\x*0,\y*5)(\x*20,\y*5)};
	\draw [very thin,color=gray] plot [smooth] coordinates {(\x*0,\y*10)(\x*20,\y*10)};
	\draw [very thin,color=gray] plot [smooth] coordinates {(\x*0,\y*15)(\x*20,\y*15)};
	\draw [very thin,color=gray] plot [smooth] coordinates {(\x*0,\y*20)(\x*20,\y*20)};
	\draw [very thin,color=gray] plot [smooth] coordinates {(\x*0,\y*25)(\x*20,\y*25)};
	\draw [very thin,color=gray] plot [smooth] coordinates {(\x*0,\y*30)(\x*20,\y*30)};
	\draw [very thin,color=gray] plot [smooth] coordinates {(\x*0,\y*35)(\x*20,\y*35)};
	\draw [very thin,color=gray] plot [smooth] coordinates {(\x*0,\y*0)(\x*0,\y*35)};
	\draw [very thin,color=gray] plot [smooth] coordinates {(\x*5,\y*0)(\x*5,\y*35)};
	\draw [very thin,color=gray] plot [smooth] coordinates {(\x*10,\y*0)(\x*10,\y*35)};
	\draw [very thin,color=gray] plot [smooth] coordinates {(\x*15,\y*0)(\x*15,\y*35)};
	\draw [very thin,color=gray] plot [smooth] coordinates {(\x*20,\y*0)(\x*20,\y*35)};
	\draw [very thin,color=gray] plot [smooth] coordinates {(\x*25,\y*0)(\x*45,\y*0)};
	\draw [very thin,color=gray] plot [smooth] coordinates {(\x*25,\y*5)(\x*45,\y*5)};
	\draw [very thin,color=gray] plot [smooth] coordinates {(\x*25,\y*10)(\x*45,\y*10)};
	\draw [very thin,color=gray] plot [smooth] coordinates {(\x*25,\y*15)(\x*45,\y*15)};
	\draw [very thin,color=gray] plot [smooth] coordinates {(\x*25,\y*20)(\x*45,\y*20)};
	\draw [very thin,color=gray] plot [smooth] coordinates {(\x*25,\y*25)(\x*45,\y*25)};
	\draw [very thin,color=gray] plot [smooth] coordinates {(\x*25,\y*30)(\x*45,\y*30)};
	\draw [very thin,color=gray] plot [smooth] coordinates {(\x*25,\y*35)(\x*45,\y*35)};
	\draw [very thin,color=gray] plot [smooth] coordinates {(\x*25,\y*0)(\x*25,\y*35)};
	\draw [very thin,color=gray] plot [smooth] coordinates {(\x*30,\y*0)(\x*30,\y*35)};
	\draw [very thin,color=gray] plot [smooth] coordinates {(\x*35,\y*0)(\x*35,\y*35)};
	\draw [very thin,color=gray] plot [smooth] coordinates {(\x*40,\y*0)(\x*40,\y*35)};
	\draw [very thin,color=gray] plot [smooth] coordinates {(\x*45,\y*0)(\x*45,\y*35)};
		
	\draw [postaction={draw,decorate,decoration={shape backgrounds,shape=circle,shape size=1.25mm,%
        shape sep={0.183cm, between center}},fill=black}] plot [thick,smooth] coordinates {(\x*25,\y*15)(\x*35,\y*15)};
    \draw [postaction={draw,decorate,decoration={shape backgrounds,shape=circle,shape size=1.25mm,%
        shape sep={0.183cm, between center}},fill=black}] plot [thick,smooth] coordinates {(\x*36,\y*26)(\x*45,\y*26)};
   	\draw [postaction={draw,decorate,decoration={shape backgrounds,shape=circle,shape size=1.25mm,%
        shape sep={0.183cm, between center}},fill=black}] plot [thick,smooth] coordinates {(\x*0,\y*10)(\x*5,\y*10)};
    \draw [postaction={draw,decorate,decoration={shape backgrounds,shape=circle,shape size=1.25mm,%
        shape sep={0.183cm, between center}},fill=black}] plot [thick,smooth] coordinates {(\x*6,\y*21)(\x*16,\y*21)};
    \draw [postaction={draw,decorate,decoration={shape backgrounds,shape=circle,shape size=1.25mm,%
        shape sep={0.183cm, between center}},fill=black}] plot [thick,smooth] coordinates {(\x*17,\y*32)(\x*20,\y*32)};
	
\end{tikzpicture}
\end{subfigure}
\caption{The $x$-axis represents the discrete times $k^\prime\in\mathbb{N}$, while the $y$-axis shows set-valued maps $K_1^1$ and $K_1^2$. Note that $K_1^1(k^\prime)$ and $K_1^2(k^\prime)$ map from a natural number to a set of natural numbers.}
\label{fig:fig_k1calc}
\end{figure*}
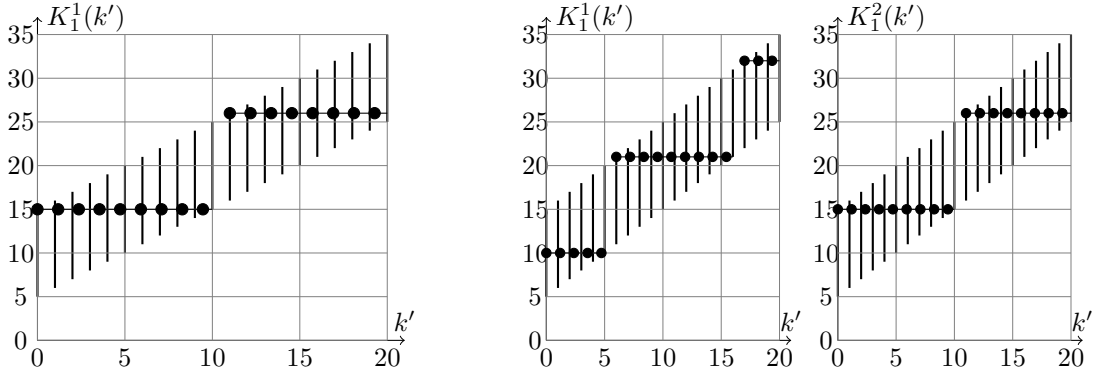

For all-time satisfying formulas $\phi:=G_{[0,\infty ]}\psi$, we need $(\boldsymbol{x},k^\prime)\models \psi$ for all $k^\prime\in\mathbb{N}$. We hence have to find  $k_1^i$ for each $k^\prime\in\mathbb{N}$ and each $i\in\{1,\hdots,N^\phi\}$. Therefore, we consider the functions $k_1^i:\mathbb{N}\to\mathbb{N}$ so that $k_1^i(k^\prime)$ dictates which $k_1^i$ is used at $k^\prime$. 

\begin{example}\label{example55}
Let $\phi_1:=G_{[0,\infty ]}\psi_1$ with $\psi_1:=F_{\{5,\hdots,15\}}\gamma_1$ and $N^{\phi_1}=1$. The set-valued map $K_1^1:\mathbb{N}\rightrightarrows\mathbb{N}$ is defined by $K_1^1(k^\prime)\in\{k^\prime+5,k^\prime+15\}$. It holds that $(\boldsymbol{x},k^\prime)\models \psi_1$ if and only if $(\boldsymbol{x},k_1^1(k^\prime))\models \gamma_1$ with $k_1^1(k^\prime)\in K_1^1(k^\prime)$. In Fig. \ref{fig:t1_precalc_idea} for instance, a vertical line at $k^\prime=5$, i.e., $K_1^1(5)$, indicates that $\gamma_1$ should be true at least once between $\{10,11,\hdots,20\}$ to obtain $(\boldsymbol{x},5)\models \psi_1$, i.e., if there exists $k_1^1\in\{10,11,\hdots,20\}$ such that $(\boldsymbol{x},k_1^1)\models\gamma_1$, then $(\boldsymbol{x},5)\models \psi_1$. By looking at Fig. \ref{fig:t1_precalc_idea}, it turns out that if $(\boldsymbol{x},k_1^1(k^\prime))\models\gamma_1$ with $k_1^1(k^\prime):=15$ (first dotted horizontal line) for all $k^\prime \in \{0,\hdots,10\}$, then $(\boldsymbol{x},k^\prime)\models \psi_1$ for all $k^\prime \in \{0,\hdots,10\}$. If further $(\boldsymbol{x},k_1^1(k^\prime))\models\gamma_1$ with $k_1^1(k^\prime):=26$ (second dotted horizontal line) for all $k^\prime \in \{11,\hdots,20\}$, then $(\boldsymbol{x},k^\prime)\models \psi_1$ for all $k^\prime \in \{11,\hdots,20\}$. Define the function $k_1^1(k^\prime):=k_0^1 + \Delta\big\lfloor\frac{k^\prime}{\Delta}\big\rfloor$ with $k_0^1:=15$ and $\Delta:=11$, where $\lfloor x\rfloor$ rounds to the nearest integer less than or equal to $x$. If $(\boldsymbol{x},k_1^1(k^\prime))\models \gamma_1$ for all $k^\prime\in \mathbb{N}$, then $(\boldsymbol{x},k^\prime)\models \psi_1$ for all $k^\prime\in \mathbb{N}$ and consequently $(\boldsymbol{x},0)\models \phi_1$; $k_0^1$ is the point, referred to as baseline, which is periodically repeated with $\Delta$. In Fig. \ref{fig:t1_precalc_idea}, $k_1^1(k^\prime):=k_0^1 + \Delta\big\lfloor\frac{k^\prime}{\Delta}\big\rfloor$ corresponds to a periodic repetition of the dotted horizontal line at $k_0^1$ with $\Delta$. This idea can be extended in the case of two or more eventually- and until-operators, i.e., $N^\phi>1$. Fig. \ref{fig:t1_precalc_idea1} shows this situation for $\phi_2:=G_{[0,\infty]}(\psi_1 \wedge \psi_2)$ with $\psi_1$ as before, $\psi_2:=F_{\{5,\hdots,15\}}\gamma_2$, and $N^{\phi_2}=2$. Similarly, it is needed that $k_1^1(k^\prime)\in K_1^1(k^\prime)$ and $k_1^2(k^\prime)\in K_1^2(k^\prime)$. Hence, $k_1^1(k^\prime):=k_0^1 + \Delta\big\lfloor\frac{k^\prime+b_1-k_0^1}{\Delta}\big\rfloor$ with $k_0^1:=10$ and $\Delta:=11$ is chosen for the first subformula (left subfigure) and $k_1^2(k^\prime):=k_0^2 + \Delta\big\lfloor\frac{k^\prime}{\Delta}\big\rfloor$ with $k_0^2:=15$ is chosen for the second subformula (right subfigure). The distance between $k_0^1$ and $k_0^2$ is given by $\eta:=|k_0^1-k_0^2|=5$, also explained in the discussion below. If now $(\boldsymbol{x},k_1^{1}(k^\prime))\models \gamma_1$ and $(\boldsymbol{x},k_1^{2}(k^\prime))\models \gamma_2$ for all $k^\prime\in\mathbb{N}$, then $(\boldsymbol{x},k^\prime) \models \psi_1\wedge\psi_2$ for all $k^\prime \in \mathbb{N}$ and hence $(\boldsymbol{x},0)\models \phi$. 
\end{example}

Algorithm \ref{alg:1} next describes the procedure to calculate $k_1^{i}(k^\prime)$ for each $i\in\{1,\hdots,N^\phi\}$. We remark that this algorithm is not complete with respect to the feasibility of the optimization problem \eqref{eq:problem1}. In other words, if Algorithm \ref{alg:1} returns an error, it does not mean that there exists no $k_1^i$ such that \eqref{eq:problem1} is feasible. Furthermore, if Algorithm \ref{alg:1} returns the functions $k_1^i$, it does not mean that \eqref{eq:problem1} is feasible. The algorithm delivers, however, candidate $k_1^i$ that can lead to a feasible optimization problem. A lazy satisfiability modulo convex-based approach as in \cite{shoukry2017smc} that iterates between Algorithm \ref{alg:1} and \eqref{eq:problem1} may account for these drawbacks. Algorithm \ref{alg:1} is not run on all eventually- and until-operators simultaneously, but only on a subset of them that are depending on each other. In this respect, maximal dependency clusters are defined.
\begin{definition}
Consider the undirected graph $\mathcal{G} := (\mathcal{V} , \mathcal{E})$ with $\mathcal{V}:=\{1,\hdots,N^\phi\}$ as the set of nodes. There exists an edge $(v_i,v_j) \in \mathcal{E} \subseteq \mathcal{V}\times \mathcal{V}$ if and only if the predicates of the $i$-th eventually- or until-operator share at least one element of $\boldsymbol{x}$ with the $j$-th eventually- or until-operator; $\Xi \subseteq \mathcal{V}$ is a maximal dependency cluster if and only if $\forall v_i,v_j \in \Xi$ there is a path from $v_i$ to $v_j$ in $\mathcal{G}$ and  $\nexists v_k \in\mathcal{V} \setminus \Xi$ such that there is a path from $v_i$ to $v_k$ in $\mathcal{G}$.
\end{definition}


Algorithm \ref{alg:1} is now applied to each dependency cluster $\Xi_l\in \{\Xi_1,\hdots,\Xi_{N_\Xi^\phi}\}$ separately. The inputs are $\Xi_l$ and the intervals $[ a_i,b_i ]$ of each eventually- and until-operators contained in $\Xi_l$ stacked such that $\boldsymbol{a}_{\Xi_l}:=\begin{bmatrix} a_{\Xi_l(1)}, \hdots, a_{\Xi_l(|\Xi_l|)} \end{bmatrix}^T$ and $\boldsymbol{b}_{\Xi_l}:=\begin{bmatrix} b_{\Xi_l(1)}, \hdots, b_{\Xi_l(|\Xi_l|)} \end{bmatrix}^T$ (line 1), where $\Xi_l(j)$ denotes the $j$-th element of $\Xi_l$ and $|\Xi_l|$ denotes the cardinality of $\Xi_l$. The outputs are $\Delta_{\Xi_l}$ and $k_0^i$ (line 19), which can then be used to determine the function $k_1^i$ given by
\begin{align}
k_1^{i}(k^\prime):=k_0^i + \Delta_{\Xi_l}\Big\lfloor \frac{k^\prime+h_d^{b_i}-k_0^i}{\Delta_{\Xi_l}}\Big\rfloor,
\label{eq:k_1_func}
\end{align}  
where $h_d^{b_i}:=\max_{k\in \kl 0,b_i\kr} k$, i.e., the discrete formula length of the $i$-th eventually- or until-operator. Note again that $k_0^i$ is a baseline that is periodically repeated with $\Delta_{\Xi_l}$ as discussed in Example \ref{example55} (horizontal lines in Fig. \ref{fig:fig_k1calc}). Lines 2-7 determine $d_{\Xi_l}^{min}$, i.e., the shortest discrete-time interval among all eventually- and until-operators in $\Xi_l$. The period $\Delta_{\Xi_l}:=d_{\Xi_l}^{min}+1$ is calculated in line 8. If the number of operators $|\Xi_l|$ is too high and the period $\Delta_{\Xi_l}$ is too short, then the algorithm returns an error (lines 9-10). Otherwise, define the distance between two consecutive $k_0^{\Xi_l(j)}$ and $k_0^{\Xi_l(j+1)}$ to be $\eta_{\Xi_l}:=\lfloor\frac{\Delta_{\Xi_l}}{|\Xi_l|}\rfloor$ (line 12). It follows that $|\Xi_l|\eta_{\Xi_l}\le \Delta_{\Xi_l}$ and consequently each eventually- and until-operator obtains a separate slot within the period $\Delta_{\Xi_l}$, which guarantees that $k_1^i\neq k_1^j$ for all $i,j\in\Xi_l$ with $i\neq j$. In line 13, the baseline $k_0^i$ for $i=\Xi_l(1)$ is calculated as $k_0^{\Xi_l(1)}:=\min_{j\in\{1,\hdots,|\Xi_l|\}}\min_{k\in\kl \boldsymbol{a}_{\Xi_l}(j),\boldsymbol{b}_{\Xi_l}(j)\kr}k$. All $k_0^i$ with $i\in \Xi_l\setminus \Xi_l(1)$ are then calculated in lines 14-16. Note that for Example \ref{example2} we get $k_0^1:=10$ and $k_0^2:=15$, while Algorithm \ref{alg:1} outputs $k_0^1:=5$ and $k_0^2:=10$. Both solutions are valid, but the values in Example \ref{example55} have been chosen this way for illustrative reasons.
 
 \algnewcommand{\algorithmicgoto}{\textbf{go to}}%
\algnewcommand{\Goto}[1]{\algorithmicgoto~\ref{#1}}%
\begin{algorithm}
  \caption{Calculation of $\Delta_{\Xi_l}$ and $k_0^i$ for all $i\in\Xi_l$}\label{alg:1}
  \begin{algorithmic}[1]
    \Procedure{$k_1$}{$\Xi_l,\boldsymbol{a}_{\Xi_l},\boldsymbol{b}_{\Xi_l}$}
    \For{$j:=1:|\Xi_l|$ }
    	  \State $\boldsymbol{k}_{min}(j) := \min_{k\in \kl \boldsymbol{a}_{\Xi_l}(j),\boldsymbol{b}_{\Xi_l}(j) \kr} k$
    	  \State $\boldsymbol{k}_{max}(j) := \max_{k\in \kl \boldsymbol{a}_{\Xi_l}(j),\boldsymbol{b}_{\Xi_l}(j) \kr} k$
      \State $\boldsymbol{d}(j) := \boldsymbol{k}_{max}(j)-\boldsymbol{k}_{min}(j)$
    \EndFor
    \State $d_{\Xi_l}^{min}:= \min_{j\in\{1,\hdots,|\Xi_l|\}} \boldsymbol{d}(j)$
    \State $\Delta_{\Xi_l} := d_{\Xi_l}^{min}+1$
    \If{$|\Xi_l|>\Delta_{\Xi_l}$}
    		\State error() \Comment{Not feasible}
    	\Else
    		\State $\eta_{\Xi_l} := \lfloor \frac{\Delta_{\Xi_l}}{|\Xi_l|} \rfloor$
    		\State $k_0^{\Xi_l(1)}:=\min_{j\in\{1,\hdots,|\Xi_l|\}} \boldsymbol{k}_{min}(j)$
    		\For{$j:=2:|\Xi_l|$}
    			\State $k_0^{\Xi_l(j)} := k_0^{\Xi_l(1)} + (j-1)\eta_{\Xi_l}$
    		\EndFor
    \EndIf
    \State \textbf{return} $\Delta_{\Xi_l}, k_0^i $ for all $i\in\Xi_l$
  \EndProcedure
\end{algorithmic}
\end{algorithm}

\subsection{Computationally-efficient Solutions via Convex Quadratic Programming}
\label{sec:our_approach}

It is next shown how Algorithm \ref{alg:1} is used in our control approach defined in Problem \ref{problem1}. We emphasize beforehand that Algorithm \ref{alg:1} is used to obtain $\As{\psi}{x}{k}$ in \eqref{eq:cost_111} and to make sure that \eqref{eq:constraint_problem} holds. For all-time satisfying formulas, we show in a first step (Step A) how $G_{[ 0,\infty]}\psi$ with $\psi$ as 1) $\gamma_1 \until{a}{b} \gamma_2$, 2) $F_{[a,b]}\gamma_1$, and 3) $G_{[a,b]}\gamma_1$ can be handled if $\gamma_1$ and $\gamma_2$ do not contain conjunctions. If $\gamma_1$ and/or $\gamma_2$ contain a conjunction (last case in \eqref{class:mu}), e.g., $\gamma_1:=\mu_1\wedge \mu_2$ and $\gamma_2:=\mu_3\wedge \mu_4$, then it holds that $\gamma_1 \until{a}{b} \gamma_2=\mu_1 \until{a}{b} \mu_3 \wedge \mu_2 \until{a}{b} \mu_4$, $F_{[a,b]}\gamma_1=F_{[a,b]}\mu_1\wedge F_{[a,b]}\mu_2$, and $G_{[a,b]}\gamma_1=G_{[a,b]}\mu_1\wedge G_{[a,b]}\mu_2$  if $k_1$, as determined by Algorithm \ref{alg:1}, remains the same for the until- and eventually-operator on the right-hand side of the equations (see Definition \ref{def:01}). This case hence reduces to the conjunction case to be discussed in the second step (Step B) where we consider $G_{[ 0,\infty]}\psi$ with $\psi$ as $\psi_1 \wedge \psi_2$. Note that one-time satisfying formulas are a subclass of all-time satisfying formulas with $k_l=k_h=k_{event}$.

\textbf{Step A)} First, consider $\psi:=\gamma_1 \until{a}{b} \gamma_2$ with $\As{\gamma_1}{x}{k}:=z_1(k)$ and $\As{\gamma_2}{x}{k}:=z_2(k)$. For $\phi:=G_{[ 0,\infty]} \psi$, the equations \eqref{eq:cost_111} and \eqref{eq:constraint_problem} in Problem \ref{prob:gen} become
\begin{subequations}\label{opt_problem}
\begin{align}
&\hspace{-0.65cm}\underset{\boldsymbol{u}_{st}}{\operatorname{argmax}} \; \frac{1}{2} \sum_{k^\prime=k_l}^{k_h}  z_2\big(k_1(k^\prime)\big) 
 +\frac{1}{k_1(k^\prime)-k^\prime+1}    \sum_{k^{\prime\prime}=k^\prime}^{k_1(k^\prime)}z_1(k^{\prime\prime})  \label{eq:un1}\\
\text{s.t. } &z_1(k^{\prime\prime})\ge 0 \; \; \;  \forall \ k^\prime \in \kl \tau_{k_l},\tau_{k_h}\kr, \forall  k^{\prime\prime} \in \kl \tau_{k^\prime},\tau_{k_1(k^\prime)}\kr\label{eq:1111a}\\
		&z_2\big(k_1(k^\prime)\big)\ge 0 \;\;\; \forall \ k^\prime \in \kl \tau_{k_l},\tau_{k_h} \kr.\label{eq:1111b}
\end{align}
\end{subequations}
Note that the constraints \eqref{eq:1111a} and \eqref{eq:1111b} are added to account for the constraint \eqref{eq:constraint_problem}, i.e., these constraints establish the desired under-approximation property and hence lead to satisfaction of $\psi$ at each $k^\prime$. The function $k_1(k^\prime):=k_1^1(k^\prime)$ is given by \eqref{eq:k_1_func}. Second, consider $\psi:=F_{[ a,b]}\gamma_1$ with $\As{\gamma_1}{x}{k}:=z_1(k)$. For $\phi:=G_{[ 0,\infty]} \psi$, the equations \eqref{eq:cost_111} and \eqref{eq:constraint_problem} become 
\begin{subequations}\label{eq:opt_ev}
\begin{align}
&\underset{\boldsymbol{u}_{st}}{\operatorname{argmax}} \sum_{k^\prime=k_l}^{k_h} z_1\big(k_1(k^\prime)\big) \label{eq:un2}\\
\text{s.t. } 	&z_1\big(k_1(k^\prime)\big)\ge 0 \;\;\; \forall \ k^\prime \in \kl \tau_{k_l},\tau_{k_h}\kr \label{eq:1111c}
\end{align}
\end{subequations}
where the constraint \eqref{eq:1111c} enforces \eqref{eq:constraint_problem} in Problem \ref{prob:gen}. Third, consider $\psi:=G_{[ a,b]}\gamma_1$ with $\As{\gamma_1}{x}{k}:=z_1(k)$. For $\phi:=G_{[ 0,\infty]} \psi$, the equations \eqref{eq:cost_111} and \eqref{eq:constraint_problem} become
\begin{subequations}\label{eq:opt_al}
\begin{align}
&\underset{\boldsymbol{u}_{st}}{\operatorname{argmax}} \sum_{k^\prime=k_l}^{k_h} \frac{1}{k_{max,k^\prime}-k_{min,k^\prime}+1}\sum_{k^{\prime\prime}=k_{min,k^\prime}}^{k_{max,k^\prime}} z_1(k^{\prime\prime})\label{eq:un3}\\
\begin{split}
\text{s.t. } 	&z_1(k^{\prime\prime})\ge 0  \; \; \; \forall \ k^\prime \in \kl \tau_{k_l},\tau_{k_h}\kr,\\ &\hspace{2.5cm}\forall \; k^{\prime\prime} \in \kl \tau_{k^\prime}+a,\tau_{k^\prime}+b\kr  \label{eq:1111d}
\end{split}
\end{align}
\end{subequations}
where $k_{min,k^\prime}:=\min_{k^*\in \kl \tau_{k^\prime}+a,\tau_{k^\prime}+b \kr} k^*$, $k_{max,k^\prime}:=\max_{k^*\in \kl \tau_{k^\prime}+a,\tau_{k^\prime}+b\kr} k^*$, and the constraint \eqref{eq:1111d} is again added to enforce \eqref{eq:constraint_problem} in Problem \ref{prob:gen}. 

\textbf{Step B)} Consider $\psi := \psi_i \wedge \psi_j$ where $\psi_i$, $\psi_j$ are of class $\psi$ as in \eqref{class:psi} and do not contain conjunctions/disjunctions. For $\phi:=G_{[ 0,\infty]} \psi$, the equations \eqref{eq:cost_111} and \eqref{eq:constraint_problem} become
\begin{subequations}\label{eq:conj_opt}
\begin{align}
&\underset{\boldsymbol{u}_{st}}{\operatorname{argmax}} \sum_{k^\prime=k_l}^{k_h} \text{min}(\As{\psi_i}{x}{k^\prime},\As{\psi_j}{x}{k^\prime})\label{eq:conj_opt111}\\
\text{s.t. } &\At{\psi_i} \text{ and } \At{\psi_j} ,
\end{align}
\end{subequations}
where $\At{\psi_i}$ is a shorthand notation for the constraints \eqref{eq:1111a} and \eqref{eq:1111b} if $\psi_i=\gamma_1 \until{a}{b} \gamma_2$, \eqref{eq:1111c} if $\psi_i=F_{[ a,b]}\gamma_1$ and \eqref{eq:1111d} if $\psi_i=G_{[ a,b]}\gamma_1$. 

\begin{thm}\label{theorem:1}
Assume that $\boldsymbol{z}(k)$ is linear as in \eqref{eq:stacked_pred}, $\psi$ does not contain disjunctions, and that \eqref{opt_problem}, \eqref{eq:opt_ev}, \eqref{eq:opt_al}, and \eqref{eq:conj_opt} together with Algorithm \ref{alg:1} are used to encode \eqref{eq:problem1}. Then the optimization problem \eqref{eq:problem1} can be written as a convex quadratic program. If $M:=\underbar{0}_{m,m}$, then the optimization problem can be written as a linear program. 
\end{thm}
\begin{pf}
Equations \eqref{opt_problem}, \eqref{eq:opt_ev}, \eqref{eq:opt_al} can be reduced to a convex quadratic program as $\underset{\boldsymbol{u}_{st}}{\operatorname{argmax}} \;  \boldsymbol{1}_N^T E\; \boldsymbol{z}_{all}$ s.t. $ R\boldsymbol{z}_{all}\ge \boldsymbol{0}$ where $\boldsymbol{z}_{all}$ stacks $\boldsymbol{z}_{st}$ with past predicates as $\boldsymbol{z}_{all}:=\begin{bmatrix}
{\boldsymbol{z}(k_l)}^T&
\hdots&
{\boldsymbol{z}(k_0)}^T&
\boldsymbol{z}^T_{st}
\end{bmatrix}^T$. Note that $\boldsymbol{z}_{all}$ is linear in $\boldsymbol{u}_{st}$ as can be seen from \eqref{eq:transformation} and that a linear program is obtained when $M:= \underbar{0}_{m,m}$. For translating \eqref{opt_problem} into $E$ and $R$, note that $E$ depends on $k_1(k^\prime)$ and is of size $E\in\mathbb{R}^{N\times N_\mu (N+h_d^\psi)}$ where $N_\mu=1$. By denoting $i_k:=i+k_l-1$, the $E$ matrix can be constructed as
\begin{align}
E(i,j):=\begin{cases}
\frac{1}{2} \hspace{1.9cm}\text{if:}  \; j=2(h_d^\psi+k_1(i_k)-k_0)\\
\frac{1}{2(k_1(i_k)-i_k+1)} \hspace{0.1cm}\text{if:}\; 1)  \hspace{0.035cm}j \text{ is odd, and} \\
\hspace{2.53cm} 2) \hspace{0.035cm} 2(i-1)\le j, \text{ and}\\
\hspace{2.53cm} 3) \hspace{0.035cm} j\le 2(h_d^\psi+k_1(i_k)-k_0)  \\
0 \hspace{1.9cm}\text{ otherwise.}
\end{cases}
\label{eq:E_until}
\end{align}
Let $N_c$ denote the total number of inequalities in \eqref{eq:1111a} and \eqref{eq:1111b}. Then the matrix $R\in\mathbb{R}^{N_c\times N_\mu(N+h_d^\psi)}$ is selected such that $R(i,j)\boldsymbol{z}_{all}\ge \boldsymbol{0}_{N_c}$ represents all inequalities in \eqref{eq:1111a} and \eqref{eq:1111b}. Next, note that \eqref{eq:opt_ev} and \eqref{eq:opt_al} can be translated into $E$ with $N_\mu=1$, while $R$ can be determined in the same way as for $\psi:=\gamma_1 \until{a}{b} \gamma_2$. The matrix $E$ for \eqref{eq:opt_ev} is then
\begin{align}
E(i,j):=\begin{cases}
1 \hspace{0.5cm} \text{ if } \; j=h_d^\psi+k_1(i_k)-k_0\\
0 \hspace{0.5cm} \text{ otherwise,}
\end{cases}
\label{eq:E_eventually}
\end{align}
while the matrix  $E$ for \eqref{eq:opt_al} is 
\begin{align}
E(i,j):=\begin{cases}
\frac{1}{\kappa} \hspace{0.2cm}\text{if:}\; 1) \; j\le k_{max,i_k}-k_l+1, \text{ and} \\
\hspace{0.85cm} 2) \; k_{min,i_k}-k_l+1\le j\\
0 \hspace{0.275cm} \text{otherwise}
\end{cases}
\label{eq:E_always}
\end{align}
where $\kappa:=k_{max,i_k}-k_{min,i_k}+1$. For conjunctions in \eqref{eq:conj_opt}, note that $\underset{\boldsymbol{u}_{st}}{\operatorname{argmax}}$ $\sum_{k^\prime=k_l}^{k_h} \text{min}(\As{\psi_i}{x}{k^\prime},\As{\psi_j}{x}{k^\prime})$ in \eqref{eq:conj_opt111} is a sum of finite elements as
\begin{align}
\begin{split}
\underset{\boldsymbol{u}_{st}}{\operatorname{argmax}} \{\min(\As{\psi_i}{x}{k_l},\As{\psi_j}{x}{k_l}) 
+\hdots+\\
\min(\As{\psi_i}{x}{k_h},\As{\psi_j}{x}{k_h})\}\label{eq:eq_conj}.
\end{split}
\end{align}
With $\boldsymbol{u}_{\boldsymbol{x}}:=\begin{bmatrix}
u_{\boldsymbol{x},1} &
\hdots&
u_{\boldsymbol{x},N}&
\boldsymbol{u}_{st}^T
\end{bmatrix}^T
$ where $u_{\boldsymbol{x},1},\hdots,u_{\boldsymbol{x},N}$ are new decision variables, \eqref{eq:conj_opt} can hence be written as
\begin{subequations}\label{eq:conj_ref}
\begin{align}
&\hspace{-0.05cm}\underset{\boldsymbol{u}_{\boldsymbol{x}}}{\operatorname{argmax}} \; \sum_{i=1}^{N}u_{\boldsymbol{x},i} \\
\hspace{-0.3cm}\text{s.t. } &\hspace{-0.05cm}u_{\boldsymbol{x},1} \le  \As{\psi_i}{x}{k_l} \text{ and } u_{\boldsymbol{x},1} \le  \As{\psi_j}{x}{k_l} \label{eq:conj_ref111a}\\
& \vdots\\
&\hspace{-0.05cm}u_{\boldsymbol{x},N} \le  \As{\psi_i}{x}{k_h} \text{ and } u_{\boldsymbol{x},N} \le  \As{\psi_j}{x}{k_h} \label{eq:conj_ref111b}\\
&\hspace{-0.05cm}\At{\psi_i} \text{ and } \At{\psi_j}.
\end{align}
\end{subequations}
Note that \eqref{eq:conj_ref} and \eqref{eq:conj_opt} are equivalent \cite[pp. 6-7]{boyd}. This is again a convex quadratic program with  cost function
$\underset{\boldsymbol{u}_{\boldsymbol{x}}}{\operatorname{argmax}}\; \begin{bmatrix}
\boldsymbol{1}^T_N &
\boldsymbol{0}^T_{N m}
\end{bmatrix}\boldsymbol{u}_x$. By defining
$
H_{2,man} := \begin{bmatrix} 
\underline{0}_{N_\mu N,N} & H_2
\end{bmatrix}
$, the stacked predicate vector from \eqref{eq:transformation} can be rewritten as
$
\boldsymbol{z}_{st}:=H_1\boldsymbol{x}(k_0)+H_{2,man}\boldsymbol{u}_x+\boldsymbol{1}_N\otimes \boldsymbol{c}
$
so that the linear inequalities of (\ref{eq:conj_ref111a}) - (\ref{eq:conj_ref111b}) are equivalent to
$
Q\boldsymbol{u}_x \le R\boldsymbol{z}_{all}
$, where $Q$ and $R$ are constructed accordingly. Explicit construction rules for $Q$ and $R$ are omitted due to space limitation. Note, however, that the $E$ matrices constructed in \eqref{eq:E_until}, \eqref{eq:E_eventually}, and \eqref{eq:E_always} can be used to form $R$. An extension to more than one conjunction $\psi:=\psi_i \wedge \psi_j \wedge \psi_k \wedge \cdots$ can be handled by adding additional constraints for each added conjunction.  \hfill\ensuremath{\blacksquare}
\end{pf}


Convex quadratic programs can be solved reliably and efficiently. We therefore propose in Step C) to solve disjunctions by calculating multiple convex quadratic programs in parallel instead of using computationally demanding mixed integer programs to resolve the disjunction operator, e.g., by using the Big-M method. 

\textbf{Step C)} Consider next $\psi := \psi_i \vee \psi_j$ where $\psi_i$ and $\psi_j$ are of class $\psi$ as in \eqref{class:psi} and do not contain any other disjunctions for now. For $\phi:=G_{[ 0,\infty)} \psi$, we separately calculate the optimal solution of $G_{[ 0,\infty)} \psi_i$ and $G_{[ 0,\infty)} \psi_j$ with $k_l$ and $k_h$ based on $\psi$. Hence, calculate $C_\pi:= \underset{\boldsymbol{u}_{st}}{\operatorname{max}} \; \sum_{k^\prime=k_l}^{k_h} \As{\psi_\pi}{x}{k^\prime}$ 
s.t. $\At{\psi_\pi}$ for each $\pi\in\{i,j\}$ where $\At{\psi_\pi}$ denotes the corresponding set of constraints obtained from the previous steps. These two processes are run simultaneously to then choose the corresponding input with the highest cost to be applied to the system. In other words, let $\pi^*:=\text{argmax}_{\pi\in\{i,j\}} C_\pi$ and apply the corresponding input $\boldsymbol{u}_{st}^*$ to \eqref{system_discrete} where $\boldsymbol{u}_{st}^*$ is calculated as $\boldsymbol{u}_{st}^*:= \underset{\boldsymbol{u}_{st}}{\operatorname{argmax}} \; \sum_{k^\prime=k_l}^{k_h} \As{\psi_{\pi^*}}{x}{k^\prime}$ s.t. $\At{\psi_{\pi^*}}$. This procedure can be applied in exactly the same way to solve formulas $\psi$ involving more than one disjunction, where additional disjunctions lead to additional $C_\pi$'s. Due to the choice of $k_l$ and $k_h$ based on $\psi$, a receding horizon procedure will guarantee satisfaction of $\phi$ in the sense of Corollary \ref{cor3}. It could be argued that this way of handling disjunctions is rather conservative since we assume that either $\psi_i$ or $\psi_j$ holds for each $k^\prime\in\{k_l,\hdots,k_h\}$. There may indeed be cases where $(\boldsymbol{x},k^\prime)\models\psi_i$ for some $k^\prime$ and $(\boldsymbol{x},k^\prime)\models\psi_j$ for the remaining $k^\prime$ yield a feasible result. Our approach, however, is computationally more favorable than solving mixed integer programs as will be further argued in the simulation results of Section \ref{sec:case_study}.

We now analyze how the framework derived above can be modified when the optimization problem is infeasible, i.e., when there is no solution to \eqref{eq:problem1}. The idea is similar to \cite{sadraddini} and \cite{ghosh2016diagnosis} and makes use of slack variables $\zeta_i\ge 0$ with $i\in\{1,\hdots,N_\mu\}$ that are added as decision variables to the optimization problem. Therefore, recall that $\boldsymbol{\zeta}:=\begin{bmatrix} \zeta_1 & \zeta_2 & \hdots & \zeta_{N_\mu}\end{bmatrix}^T$ and define $\boldsymbol{u}_{sl}:=\begin{bmatrix} \boldsymbol{u}_{st}^T & \boldsymbol{\zeta}^T \end{bmatrix}^T$. Next, change the cost function in \eqref{eq:cost_111} to 
\begin{align}\label{eq:modified_cost}
\begin{split}
&\underset{\boldsymbol{u}_{sl}}{\operatorname{argmax}} \sum_{k^\prime=k_l}^{k_h} \Asxi{\psi}{x}{k^\prime}-s\boldsymbol{1}_{N_\mu}^T\boldsymbol{\zeta},
\end{split}
\end{align}
 where $s$ is a sufficiently large constant. Recall the difference between $\Asxi{\psi}{x}{k^\prime}$ and $\As{\psi}{x}{k^\prime}$ as discussed in Section \ref{sec:robustness_notions}, i.e., the former is evaluated with $\boldsymbol{z}_{\boldsymbol{\zeta}}(k)$ and the latter is evaluated with $\boldsymbol{z}(k)$. This leads to the fact that the constraints in \eqref{eq:1111a}, \eqref{eq:1111b}, \eqref{eq:1111c}, and \eqref{eq:1111d} are changed to 
$z_1(k^{\prime\prime}) + \zeta_1 \ge 0 $, 
$z_2\big(k_1(k^\prime)\big)+ \zeta_2 \ge 0$, 
$z_1\big(k_1(k^\prime)\big) + \zeta_1 \ge 0$, and
$z_1(k^{\prime\prime}) + \zeta_1 \ge 0 $, respectively. 

\begin{corollary}\label{relaxation11}
If the optimization problem \eqref{eq:problem1} is infeasible, then the modified cost function in \eqref{eq:modified_cost} together with the use of $\boldsymbol{z}_{\boldsymbol{\zeta}}(k)$ leads to a least violating solution of the optimization problem in \eqref{eq:problem1}.
\end{corollary}
\begin{pf}
Note that formulas $\psi$ in \eqref{class:psi} are in PNF. Then, according to Corollary \ref{cor1}, we have that $\As{\psi}{x}{k}<\Asxi{\psi}{x}{k}$ and $\Asxi{\psi}{x}{k}-\As{\psi}{x}{k}\ge \zeta_{min}$ where $\zeta_{min}:= \min_{i\in \{1,\hdots,N_\mu \}} \zeta_i$ if $\zeta_i>0$ for all $i\in\{1,\hdots,N_\mu\}$. The cost function minimizes $\zeta_i$, but will select $\zeta_i>0$ if the original problem ($\boldsymbol{\zeta}=\boldsymbol{0}_{N_\mu}$) is infeasible in order to artificially increase the robustness $\Asxi{\psi}{x}{k}$. This will eventually lead to a feasible problem, but now using $\Asxi{\psi}{x}{k}$ instead of $\As{\psi}{x}{k}$. The minimization of $s\boldsymbol{1}_{N_\mu}^T\boldsymbol{\zeta}$ leads to a least violating solution of the optimization problem in \eqref{eq:problem1}. Note that the hyperplane interpretation of Section \ref{sec:robustness_notions} applies due to the assumption of linear predicate functions.  \hfill\ensuremath{\blacksquare}
\end{pf}
\begin{remark}
Corollary \ref{relaxation11} does not guarantee that $(\boldsymbol{x},k) \models \psi \text{ for all } k\in\kl \tau_{k_l},\tau_{k_h}\kr$. We remark that \cite{ghosh2016diagnosis} uses a similar idea for the specific case of predicate repair. In contrast to our approach, \cite{sadraddini} uses a scalar slack variable $\xi$ and defines $\boldsymbol{z}_{soft}(k)=\boldsymbol{z}(k)+\boldsymbol{1}_{N_\mu}\zeta$. This works well for SR, but for DSASR a vector $\boldsymbol{\zeta}$ is more suitable since it avoids an unnecessary increase in DSASR. For instance, consider $x_1(k)=x_2(k):=0$ for all $k\in\mathbb{N}$. The formula $\phi:=(x_1\ge 5) \until{5}{10} (x_1\ge 0)$ with $\boldsymbol{z}(k):=\begin{bmatrix} x_1(k)-5 & x_2(k) \end{bmatrix}^T$ is initially infeasible due to $x_1(k)<5$, in fact we have $\rss{\phi}{0}=-5$ and $\As{\phi}{x}{0}=-5$. If instead $\boldsymbol{z}_{soft}(k):=\boldsymbol{z}(k)+\boldsymbol{1}_{N_\mu}\zeta=\begin{bmatrix} x_1(k)-5 & x_2(k) \end{bmatrix}^T+\begin{bmatrix}5 & 5 \end{bmatrix}^T$ is used, the problem is feasible with $\rsxis{\phi}=0$ and $\Asxis{\phi}{x}{k}=2.5$. For $\boldsymbol{z}_{\boldsymbol{\zeta}}(k):=\boldsymbol{z}(k)+\boldsymbol{\zeta}=\begin{bmatrix} x_1(k)-5 & x_2(k)\end{bmatrix}^T+\begin{bmatrix} 5 & 0\end{bmatrix}^T$ we get $\rsxi{\phi}=0$, while $\Asxi{\phi}{x}{k}=0$, i.e., $\Asxi{\phi}{x}{k}$ does not unnecessarily increase and indicate too optimistic robustness values.
\end{remark}

\section{Simulations}
\label{sec:case_study}
The system under consideration is a centralized multi-agent system with three agents $\alpha_1$, $\alpha_2$, and $\alpha_3$. Each agent $\alpha_i$ with $i\in\{1,2,3\}$ has four states $x^i$, $y^i$, $v_x^i$, and $v_y^i$ denoting position and velocity in two-dimensional space, respectively. The agents obey continuous-time double integrator dynamics and the stacked vector of all agents is denoted by $\boldsymbol{x}:=\begin{bmatrix} \boldsymbol{x}^1 & \boldsymbol{x}^2 & \boldsymbol{x}^3 \end{bmatrix}^T\in\mathbb{R}^{12}$ with $\boldsymbol{x}^i:=\begin{bmatrix}x^i & v_x^i & y^i & v_y^i \end{bmatrix}^T\in\mathbb{R}^4$. After periodic sampling with $T:=0.1$, the discrete-time dynamics are  
\begin{align}\label{eq:sim_dynamics}
\boldsymbol{x}(k+1) = A\boldsymbol{x}(k)+
B\boldsymbol{u}(k),
\end{align}
where $A:=\mathcal{I}_6 \otimes \begin{bmatrix}
1 & 0.1\\
0 & 1
\end{bmatrix}$, $B:=\mathcal{I}_6 \otimes \begin{bmatrix}
0.005\\
0.1
\end{bmatrix}$, and $\boldsymbol{u}(k):=\begin{bmatrix} u_x^1 & u_y^1 & u_x^2 & u_y^2 & u_x^3 & u_y^3 \end{bmatrix}^T\in [-20,20]^6$ with $\mathcal{I}_6$ being the $6\times6$ identity matrix. In order to obtain linear predicates, we use the infinity norm. For instance,  $\Big\|\begin{bmatrix} x^i & y^i \end{bmatrix}\Big\|_\infty:=\max(|x^i|,|y^i|)$ can be used to rewrite $\Big\|\begin{bmatrix} x^i & y^i \end{bmatrix}\Big\|_\infty\le 5$ as $(x^i\le 5) \wedge (-x^i\le 5) \wedge (y^i\le 5) \wedge (-y^i\le 5)$, i.e., the formula $\Big\|\begin{bmatrix} x^i & y^i \end{bmatrix}\Big\|_\infty\le 5$ can be rewritten by using three conjunctions and four predicates with the linear predicate functions $f_1(\boldsymbol{x}):=5-x^i$, $f_2(\boldsymbol{x}):=5+x^i$, $f_3(\boldsymbol{x}):=5-y^i$, and $f_4(\boldsymbol{x}):=5+y^i$.

\begin{figure*}
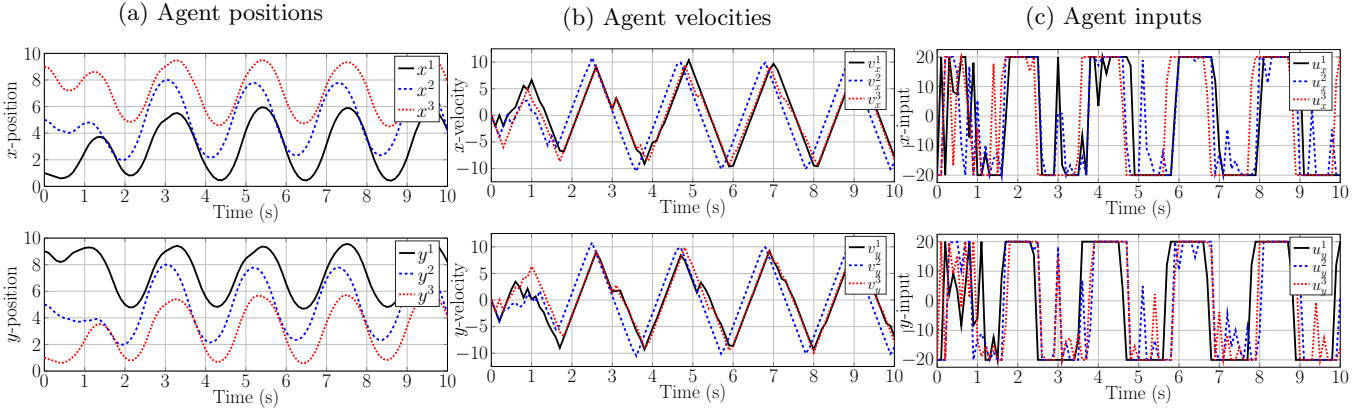

\centering
\begin{subfigure}{0.33\textwidth}\caption{Agent positions}\label{fig:1a}
\input{figures/1}
\end{subfigure}
\begin{subfigure}{0.33\textwidth}\caption{Agent velocities}\label{fig:1b}
\input{figures/2}
\end{subfigure}
\begin{subfigure}{0.33\textwidth}\caption{Agent inputs}\label{fig:1c}
%
%
\begin{tikzpicture}[scale=0.35]

\begin{axis}[%
width=6.028in,
height=1.99in,
at={(1.011in,3.406in)},
scale only axis,
xmin=0,
xmax=10,
xlabel={Time (s)},
xmajorgrids,
ymin=-22.5,
ymax=22.5,
ylabel={$x$-input},
ymajorgrids,
axis background/.style={fill=white},
legend style={legend cell align=left,align=left,draw=white!15!black},
title style={font=\LARGE},xlabel style={font=\LARGE},ylabel style={font=\LARGE},legend style={font=\Large},ticklabel style={font=\LARGE},
]
\addplot [color=black,solid,line width=2.0pt]
  table[row sep=crcr]{%
0	-20\\
0.1	20\\
0.2	-20\\
0.3	20\\
0.4	8.19444444444445\\
0.5	7.36377252921373\\
0.6	20\\
0.7	20\\
0.8	-7.26861373920196\\
0.9	18.0762349941271\\
1	-20\\
1.1	-20\\
1.2	-12.7778651994338\\
1.3	-20\\
1.4	-20\\
1.5	-20\\
1.6	-20\\
1.7	-1.6664044016987\\
1.8	20\\
1.9	20\\
2	20\\
2.1	20\\
2.2	20\\
2.3	20\\
2.4	20\\
2.5	20\\
2.6	-19.7759103641453\\
2.7	-20\\
2.8	-20\\
2.9	-20\\
3	20\\
3.1	-16.6445454114895\\
3.2	-20\\
3.3	-20\\
3.4	-20\\
3.5	-20\\
3.6	-6.65072820096318\\
3.7	-20\\
3.8	20.0000000000001\\
3.9	20\\
4	3.41533032899487\\
4.1	20\\
4.2	14.3281904308884\\
4.3	20\\
4.4	20\\
4.5	20\\
4.6	20\\
4.7	20\\
4.8	17.0700495081339\\
4.9	-20\\
5	-20\\
5.1	-20\\
5.2	-20\\
5.3	-20\\
5.4	-20\\
5.5	-20\\
5.6	-20\\
5.7	-20\\
5.8	-20\\
5.9	1.6095864409561\\
6	20\\
6.1	20\\
6.2	20\\
6.3	20\\
6.4	20\\
6.5	20\\
6.6	20\\
6.7	20\\
6.8	20\\
6.9	11.6942036978844\\
7	-11.4727053387232\\
7.1	-20\\
7.2	-20\\
7.3	-20\\
7.4	-20\\
7.5	-20\\
7.6	-20\\
7.7	-20\\
7.8	-20\\
7.9	-20\\
8	-1.3788954662377\\
8.1	20\\
8.2	20\\
8.3	20\\
8.4	20\\
8.5	20\\
8.6	20\\
8.7	20\\
8.8	20\\
8.9	18.291214097365\\
9	20\\
9.1	-20\\
9.2	-20\\
9.3	-20\\
9.4	-20\\
9.5	-20\\
9.6	-20\\
9.7	-20\\
9.8	-20\\
9.9	-20\\
10	-20\\
};
\addlegendentry{$u_x^1$};

\addplot [color=blue,dashed,line width=2.0pt]
  table[row sep=crcr]{%
0	-20\\
0.1	-20\\
0.2	20\\
0.3	20\\
0.4	20\\
0.5	14.4827586206897\\
0.6	-9.89028213166145\\
0.7	-20\\
0.8	16.7493085008298\\
0.9	-20\\
1	5.35808010714601\\
1.1	-20\\
1.2	-9.87108911450171\\
1.3	-20.0000000000001\\
1.4	-14.5330578512396\\
1.5	-12.4522252132043\\
1.6	-20\\
1.7	20\\
1.8	20\\
1.9	20\\
2	20\\
2.1	20\\
2.2	20\\
2.3	20\\
2.4	20\\
2.5	20\\
2.6	-13.8385511944522\\
2.7	-20\\
2.8	-20\\
2.9	-20\\
3	1.04314897599673\\
3.1	0.704803823064647\\
3.2	-20\\
3.3	-16.6234244685258\\
3.4	-20\\
3.5	-20\\
3.6	-14.105909651095\\
3.7	-20\\
3.8	20.0000000000001\\
3.9	20\\
4	17.1564100655464\\
4.1	20\\
4.2	20\\
4.3	20\\
4.4	20\\
4.5	20\\
4.6	20\\
4.7	-10.0714169342593\\
4.8	-20\\
4.9	-20\\
5	-19.9999999999998\\
5.1	18.9563444853246\\
5.2	-11.2065896845404\\
5.3	-14.8448111964047\\
5.4	-20\\
5.5	-20\\
5.6	-12.3850618037466\\
5.7	-20\\
5.8	-20\\
5.9	-0.9615586287855\\
6	20\\
6.1	20\\
6.2	20\\
6.3	20\\
6.4	18.6498438727048\\
6.5	20\\
6.6	20\\
6.7	20\\
6.8	20\\
6.9	-19.0452273662161\\
7	-13.087911083306\\
7.1	-20\\
7.2	-4.47282848198023\\
7.3	-17.858563872754\\
7.4	-15.1575065253675\\
7.5	-19.164547701012\\
7.6	-20\\
7.7	-20\\
7.8	-20\\
7.9	-17.0876256842356\\
8	20\\
8.1	19.0778412888562\\
8.2	11.1962071128961\\
8.3	20\\
8.4	20\\
8.5	20\\
8.6	20\\
8.7	20\\
8.8	20\\
8.9	17.0908862019535\\
9	-20\\
9.1	-20\\
9.2	-20\\
9.3	2.33618233513231\\
9.4	-20\\
9.5	-20\\
9.6	-20\\
9.7	-20\\
9.8	-4.57829459866422\\
9.9	-20\\
10	-20\\
};
\addlegendentry{$u_x^2$};

\addplot [color=red,dotted,line width=2.0pt]
  table[row sep=crcr]{%
0	-20\\
0.1	-20\\
0.2	20\\
0.3	20\\
0.4	-16.7777777777778\\
0.5	13.7969348659004\\
0.6	20\\
0.7	10.198275862069\\
0.8	0.142857142857185\\
0.9	-20\\
1	-4.11803214933892\\
1.1	-20\\
1.2	-20\\
1.3	-20\\
1.4	17.7018816197581\\
1.5	-20\\
1.6	12.1827574546657\\
1.7	20\\
1.8	20\\
1.9	20\\
2	20\\
2.1	20\\
2.2	20\\
2.3	15.1893892028713\\
2.4	20\\
2.5	-20\\
2.6	-20\\
2.7	-20\\
2.8	-20\\
2.9	-20\\
3	-20\\
3.1	-20\\
3.2	-20\\
3.3	-20\\
3.4	-20\\
3.5	-13.8051574269414\\
3.6	19.9999999999999\\
3.7	20.0000000000001\\
3.8	20\\
3.9	20\\
4	20\\
4.1	20\\
4.2	20\\
4.3	20\\
4.4	20\\
4.5	20\\
4.6	4.44814722403584\\
4.7	-20\\
4.8	-20\\
4.9	-20\\
5	-20\\
5.1	-20\\
5.2	-20\\
5.3	-20\\
5.4	-20\\
5.5	-20\\
5.6	-20\\
5.7	12.1855203619908\\
5.8	20.0000000000003\\
5.9	19.9999999999999\\
6	20\\
6.1	20\\
6.2	20\\
6.3	20\\
6.4	20\\
6.5	20\\
6.6	20\\
6.7	7.81447963800903\\
6.8	-20\\
6.9	-19.9999999999998\\
7	-20\\
7.1	-20\\
7.2	-20\\
7.3	-20\\
7.4	-20\\
7.5	-20\\
7.6	-20\\
7.7	-20\\
7.8	12.1855203619906\\
7.9	20\\
8	20\\
8.1	19.958025168109\\
8.2	20\\
8.3	20\\
8.4	20\\
8.5	20\\
8.6	20\\
8.7	20\\
8.8	7.88316572655816\\
8.9	-19.9999999999998\\
9	-20\\
9.1	-20\\
9.2	-20\\
9.3	-20\\
9.4	-20\\
9.5	-20\\
9.6	-20\\
9.7	-20\\
9.8	-20\\
9.9	12.1588091053328\\
10	20.0000000000001\\
};
\addlegendentry{$u_x^3$};

\end{axis}

\begin{axis}[%
width=6.028in,
height=1.99in,
at={(1.011in,0.642in)},
scale only axis,
xmin=0,
xmax=10,
xlabel={Time (s)},
xmajorgrids,
ymin=-22.5,
ymax=22.5,
ylabel={$y$-input},
ymajorgrids,
axis background/.style={fill=white},
legend style={legend cell align=left,align=left,draw=white!15!black},
title style={font=\LARGE},xlabel style={font=\LARGE},ylabel style={font=\LARGE},legend style={font=\Large},ticklabel style={font=\LARGE},
]
\addplot [color=black,solid,line width=2.0pt]
  table[row sep=crcr]{%
0	-20\\
0.1	-20\\
0.2	20\\
0.3	2.20529470529467\\
0.4	9.26533144275083\\
0.5	3.44827586206898\\
0.6	-8.42752227042856\\
0.7	4.66130647338306\\
0.8	20\\
0.9	-8.29836829836846\\
1	-6.03809611932988\\
1.1	20\\
1.2	-20\\
1.3	-20\\
1.4	-12.3493896358793\\
1.5	-20\\
1.6	7.97761228495309\\
1.7	20\\
1.8	20\\
1.9	20\\
2	20\\
2.1	20\\
2.2	20\\
2.3	20\\
2.4	15.5555555555556\\
2.5	-20\\
2.6	-20\\
2.7	-20\\
2.8	-20\\
2.9	-20\\
3	-20\\
3.1	-20\\
3.2	-20\\
3.3	-20\\
3.4	-20\\
3.5	-13.3307280954339\\
3.6	20\\
3.7	20\\
3.8	20\\
3.9	20\\
4	20\\
4.1	20\\
4.2	20\\
4.3	20\\
4.4	20\\
4.5	20\\
4.6	4.29000411353357\\
4.7	-20\\
4.8	-20\\
4.9	-20\\
5	-20\\
5.1	-20\\
5.2	-20\\
5.3	-20\\
5.4	-20\\
5.5	-20\\
5.6	-20\\
5.7	12.1855203619908\\
5.8	20\\
5.9	20\\
6	20\\
6.1	20\\
6.2	20\\
6.3	20\\
6.4	20\\
6.5	20\\
6.6	20\\
6.7	7.8144796380091\\
6.8	-19.9999999999999\\
6.9	-20\\
7	-20\\
7.1	-20\\
7.2	-20\\
7.3	-20\\
7.4	-20\\
7.5	-20\\
7.6	-20\\
7.7	-20\\
7.8	12.1855203619907\\
7.9	19.9999999999998\\
8	20\\
8.1	19.9580251681091\\
8.2	20\\
8.3	20\\
8.4	20\\
8.5	20\\
8.6	20\\
8.7	20\\
8.8	7.88316572655825\\
8.9	-19.9999999999999\\
9	-20\\
9.1	-20\\
9.2	-20\\
9.3	-20\\
9.4	-20\\
9.5	-20\\
9.6	-20\\
9.7	-20\\
9.8	-20\\
9.9	12.1588091053329\\
10	20\\
};
\addlegendentry{$u_y^1$};

\addplot [color=blue,dashed,line width=2.0pt]
  table[row sep=crcr]{%
0	-20\\
0.1	-20\\
0.2	-20\\
0.3	20\\
0.4	19.6666666666667\\
0.5	20\\
0.6	10.4444444444444\\
0.7	20\\
0.8	20\\
0.9	-20\\
1	-13.2407407407407\\
1.1	-11.1955555555556\\
1.2	-14.4890266299357\\
1.3	-20\\
1.4	-17.4938862453739\\
1.5	-20\\
1.6	-20\\
1.7	20\\
1.8	20\\
1.9	20\\
2	20\\
2.1	20\\
2.2	20\\
2.3	20\\
2.4	16.9238146554451\\
2.5	20\\
2.6	-20\\
2.7	-20\\
2.8	-20\\
2.9	-17.9086536419261\\
3	18.259145131861\\
3.1	-20\\
3.2	-14.0713422131226\\
3.3	-20\\
3.4	-20\\
3.5	-20\\
3.6	6.86883108635028\\
3.7	-20\\
3.8	-9.33181398952611\\
3.9	20\\
4	20\\
4.1	20\\
4.2	20\\
4.3	20\\
4.4	20\\
4.5	20\\
4.6	20\\
4.7	20\\
4.8	-20\\
4.9	-20\\
5	-20\\
5.1	4.90524196892896\\
5.2	-20\\
5.3	-20\\
5.4	-13.4702877072545\\
5.5	-20\\
5.6	-20\\
5.7	-20\\
5.8	-20\\
5.9	19.9999999999996\\
6	10.4434789409931\\
6.1	20\\
6.2	20\\
6.3	20\\
6.4	14.6891006296161\\
6.5	20\\
6.6	20\\
6.7	20\\
6.8	20\\
6.9	-16.9294050696815\\
7	-20\\
7.1	-20\\
7.2	-4.39072585815363\\
7.3	-15.5229320217527\\
7.4	-9.98690569092769\\
7.5	-16.3898769499828\\
7.6	-20\\
7.7	-9.01155641971261\\
7.8	-20\\
7.9	-20\\
8	-11.0038198278681\\
8.1	20\\
8.2	20\\
8.3	20\\
8.4	20\\
8.5	20\\
8.6	20\\
8.7	20\\
8.8	20\\
8.9	20\\
9	9.03403065853101\\
9.1	-20\\
9.2	-20\\
9.3	-20\\
9.4	-20\\
9.5	-20\\
9.6	-20\\
9.7	-20\\
9.8	-20\\
9.9	-20\\
10	1.07843576474093\\
};
\addlegendentry{$u_y^2$};

\addplot [color=red,dotted,line width=2.0pt]
  table[row sep=crcr]{%
0	-20\\
0.1	20\\
0.2	-20\\
0.3	19.5625\\
0.4	9.14062500000002\\
0.5	15.1562500000001\\
0.6	-2.21893382352949\\
0.7	20\\
0.8	1.1584738790621\\
0.9	20\\
1	-13.430884788586\\
1.1	-18.8329010772509\\
1.2	-15.4724844350695\\
1.3	-20\\
1.4	-14.952153110048\\
1.5	-20\\
1.6	-20\\
1.7	-8.65594619003278\\
1.8	20\\
1.9	20\\
2	20\\
2.1	19.8690909090908\\
2.2	20\\
2.3	20\\
2.4	20\\
2.5	20\\
2.6	-18.1236363636362\\
2.7	-20\\
2.8	-20\\
2.9	-20\\
3	14.9037276207537\\
3.1	-20\\
3.2	-13.5552249875449\\
3.3	-12.7686725435221\\
3.4	-20\\
3.5	-11.1560649291152\\
3.6	-20\\
3.7	-20\\
3.8	7.99734436319409\\
3.9	20\\
4	20\\
4.1	20\\
4.2	20\\
4.3	20\\
4.4	20\\
4.5	20\\
4.6	20\\
4.7	20\\
4.8	-20\\
4.9	-15.459245917254\\
5	-20\\
5.1	-20\\
5.2	-20\\
5.3	-20\\
5.4	2.21209441699424\\
5.5	-20\\
5.6	-20\\
5.7	-19.9261134769674\\
5.8	-20\\
5.9	9.66824638558038\\
6	20\\
6.1	20\\
6.2	20\\
6.3	20\\
6.4	20\\
6.5	20\\
6.6	20\\
6.7	20\\
6.8	17.5090276819442\\
6.9	-19.9999999999996\\
7	-3.90292609954398\\
7.1	-20\\
7.2	-11.6880318557188\\
7.3	-20\\
7.4	-20\\
7.5	-15.7577248982132\\
7.6	-20\\
7.7	-20\\
7.8	-20\\
7.9	-20\\
8	20\\
8.1	20\\
8.2	20\\
8.3	20\\
8.4	20\\
8.5	20\\
8.6	20\\
8.7	20\\
8.8	20\\
8.9	13.3912406315143\\
9	-20\\
9.1	-20\\
9.2	-20\\
9.3	-20\\
9.4	-0.574289135147436\\
9.5	-20\\
9.6	-12.6657657770272\\
9.7	-20\\
9.8	-20\\
9.9	-20\\
10	-16.0927798509682\\
};
\addlegendentry{$u_y^3$};

\end{axis}
\end{tikzpicture}%
\end{subfigure}
\caption{Simulations result for $\phi:=G_{[0,\infty)}(\psi_1\wedge\psi_2)$.}
\label{fig:1}
\end{figure*}

The specification imposed on the system, in words, is the following: each agent $\alpha_i$ is supposed to remain in the region $[0,10]\times[0,10]$, i.e., $\begin{bmatrix}
x^i & y^i
\end{bmatrix}^T\in[0,10]\times[0,10]$, and the multi-agent system is supposed to perform surveillance of this region. In formulas, the all-time satisfying formula $\phi:=G_{[0,\infty)}(\psi_1\wedge\psi_2)$ is imposed where $\psi_1:=G_{[0,3]}\Big(\Big\|\begin{bmatrix} x^1 & y^1 & x^2 & y^2 & x^3 & y^3 \end{bmatrix}^T-\boldsymbol{5}_6\Big\|_\infty\le 5\Big)$ and $\psi_2:=F_{[1,3]}\Big(\Big\|\begin{bmatrix} x^1 & y^1 \end{bmatrix}^T-\begin{bmatrix}
5 & 9
\end{bmatrix}^T\Big\|_\infty\le 1\Big)\wedge F_{[1,3]}\Big(\Big\|\begin{bmatrix} x^1 & y^1 \end{bmatrix}^T-\begin{bmatrix}
1 & 5
\end{bmatrix}^T\Big\|_\infty\le 1\Big)\wedge F_{[1,3]}\Big(\Big\|\begin{bmatrix} x^2 & y^2 \end{bmatrix}^T-\begin{bmatrix}
8 & 8
\end{bmatrix}^T\Big\|_\infty\le 1\Big)\wedge F_{[1,3]}\Big(\Big\|\begin{bmatrix} x^2 & y^2 \end{bmatrix}^T-\begin{bmatrix}
2 & 2
\end{bmatrix}^T\Big\|_\infty\le 1\Big)\wedge F_{[1,3]}\Big(\Big\|\begin{bmatrix} x^3 & y^3 \end{bmatrix}^T-\begin{bmatrix}
9 & 5
\end{bmatrix}^T\Big\|_\infty\le 1\Big)\wedge F_{[1,3]}\Big(\Big\|\begin{bmatrix} x^3 & y^3 \end{bmatrix}^T-\begin{bmatrix}
5 & 1
\end{bmatrix}^T\Big\|_\infty\le 1\Big)$; $\psi_1$ defines the requirement to stay within $[0,10]\times[0,10]$, while $\psi_2$ expresses the surveillance task that is to be performed every $1$ to $3$ seconds. This formulation results in $36$ predicates with $35$ conjunctions, which can be seen as a complex temporal logic formula.

For the simulation, we have selected a prediction horizon of $N:=45$, while no cost penalization of inputs has been used, i.e., $M:=\underline{0}_{6,6}$. Within the MPC procedure, we performed $100$ optimization steps. The closed-loop solution of this MPC procedure is shown in Fig. \ref{fig:1}. The agents remain within $[0,10]\times[0,10]$ all the time, while satisfying the surveillance task within the time bounds. This would result in $(\boldsymbol{x},0)\models\phi$ by repeating the MPC procedure infinitely many times. It can also be seen that the DSASR semantics have been maximized. For instance, looking at the signal $x^1$ and $y^1$, it can be seen that $x^1(2)=1$, $x^1(3)=5$, $x^1(4.1)=1.073$, $x^1(5.1)=4.944$, $x^1(6.2)=1.187$, and $x^1(7.2)=4.835$, while $y^1(2)=5$, $y^1(3)=9$, $y^1(4.1)=5.141$, $y^1(5.1)=8.707$, $y^1(6.2)=5.217$, and $y^1(7.2)=8.81$. This shows that the reachability task $F_{[1,3]}\Big(\Big\|\begin{bmatrix} x^1 & y^1 \end{bmatrix}^T-\begin{bmatrix}
5 & 9
\end{bmatrix}^T\Big\|_\infty\le 1\Big)\wedge F_{[1,3]}\Big(\Big\|\begin{bmatrix} x^1 & y^1 \end{bmatrix}^T-\begin{bmatrix}
1 & 5
\end{bmatrix}^T\Big\|_\infty\le 1\Big)$ for agent $\alpha_1$ is performed with high robustness. In fact, the robustness degree is $\mathcal{RD}_0(\boldsymbol{x},\phi)=0.324$. We compared this result with the case where we do not maximize the DSASR semantics, i.e., not accounting for \eqref{eq:cost_111} while still ensuring \eqref{eq:constraint_problem}, which results  in a marginal robustness degree of only $\mathcal{RD}_0(\boldsymbol{x},\phi)=0.005$. Furthermore, we remark that the open-loop solution, which is not presented due to space limitations, gives optimal robustness, i.e., $x^1(3)=x^1(5.1)=x^1(7.2)=5$, $x^1(2)=x^1(4.1)=x^1(6.2)=1$ and $y^1(3)=y^1(5.1)=y^1(7.2)=9$, $y^1(2)=y^1(4.1)=y^1(6.2)=5$. 

The simulation has been performed in real-time on a MacBook Air (Mid 2012) with a two-core 1.8 GHz CPU and 4 GB of RAM. The resulting linear program has $315$ decision variables and, on average, $2264.3$ constraints; it has been modeled in Yalmip \cite{lofberg2004yalmip} and solved by the commercial solver Gurobi \cite{gurobi}. The average time to solve  \eqref{eq:problem1} (where $\psi:=\psi_1\wedge\psi_2$) is $0.076$ seconds. For comparison, we tried to solve the same formula $\phi$ with the same dynamics \eqref{eq:sim_dynamics} by using the toolbox BluSTL \cite{raman2014model}, which is based on the works in \cite{raman1,raman2}; however, after $30$ minutes of calculations, the software still did not return a result. This confirms what the authors in \cite{raman1,raman2} already described, i.e., the MILP formulation becomes intractable when the problem size is large. In fact, checking the feasibility of the obtained MILP formulation is NP-hard. For the robustness-based encoding, \cite[Section~2.3]{raman2} states the complexity in terms of variables and constraints: $O(NN_\mu)+O(N|\phi|)$ continuous variables and $O(N|\phi|)$ binary variables are introduced, where $|\phi|$ is the number of operators in $\phi$. We remark that our simulation is made with low performance hardware, which indicates that our algorithms will also run efficiently on integrated microcontrollers.

\section{Discussion and Future Work}
\label{sec:diskussion_futurework}
This work introduced discrete average space robustness as quantitative semantics for signal temporal logic. Closely connected to these semantics, the predicate robustness degree was introduced as a new robustness notion of a signal with respect to a signal temporal logic formula by only looking at predicates. In an illustrative example, we motivated why these notions may be a suitable choice in control applications. A model predictive control framework was then presented where a simplified version of discrete average space robustness was incorporated into the cost function of the optimization problem. The presented framework guarantees the satisfaction of the formula while maximizing the discrete average space robustness. Furthermore, it was shown that the proposed framework is computationally-efficient, which was demonstrated in a high-dimensional simulation example.

For the simplified version of discrete average space robustness, a parameter $k_1$ needs to be determined. We proposed an algorithm that solves this problem; however, this algorithm has some limitations and is hence subject to future work. Another topic of future work is the question if recursive feasibility of the MPC framework can be guaranteed. Furthermore, experiments are planned in order to demonstrate the suitability of our method for practical use. In a complementary approach, we are currently also working on continuous-time feedback control laws for signal temporal logic specifications by considering nonlinear predicates and nonlinear (multi-agent) systems.






\bibliographystyle{plain}        
\bibliography{literature}           

\end{document}